\documentclass[10pt,journal,compsoc]{IEEEtran}
\ifCLASSOPTIONcompsoc
  \usepackage[nocompress]{cite}
\else
  \usepackage{cite}
\fi
\ifCLASSINFOpdf
\else
\fi

\usepackage[utf8]{inputenc} 
\usepackage[T1]{fontenc}    
\usepackage{hyperref}     
\usepackage{url}            
\usepackage[export]{adjustbox}
\usepackage{subfig}
\usepackage[misc]{ifsym}
\usepackage{epsfig}
\usepackage{graphicx}
\usepackage{algorithm,algpseudocode}
\usepackage{subfig}
\usepackage{enumitem}
\usepackage{array}
\usepackage{ragged2e}
\usepackage{color}
\usepackage{tikz}
\usepackage[edges]{forest}
\definecolor{hidden-draw}{RGB}{20,68,106}
\definecolor{hidden-pink}{RGB}{255,245,247}

\begin{document}
\title{When Search Engine Services meet Large Language Models: Visions and Challenges\vspace{-3mm}}

\author{Haoyi Xiong, \emph{Senior Member, IEEE}, Jiang Bian, \emph{Member, IEEE}, Yuchen Li, Xuhong Li, Mengnan Du, \emph{Member, IEEE}, Shuaiqiang Wang, Dawei Yin, \emph{Senior Member, IEEE}, and Sumi Helal, \emph{Fellow, IEEE}\vspace{-3mm}}


\IEEEtitleabstractindextext{%
\begin{abstract}
Combining Large Language Models (LLMs) with search engine services marks a significant shift in the field of services computing, opening up new possibilities to enhance how we search for and retrieve information, understand content, and interact with internet services. This paper conducts an in-depth examination of how integrating LLMs with search engines can mutually benefit both technologies. We focus on two main areas: using search engines to improve LLMs (Search4LLM) and enhancing search engine functions using LLMs (LLM4Search). For Search4LLM, we investigate how search engines can provide diverse high-quality datasets for pre-training of LLMs, how they can use the most relevant documents to help LLMs learn to answer queries more accurately, how training LLMs with Learning-To-Rank (LTR) tasks can enhance their ability to respond with greater precision, and how incorporating recent search results can make LLM-generated content more accurate and current. In terms of LLM4Search, we examine how LLMs can be used to summarize content for better indexing by search engines, improve query outcomes through optimization, enhance the ranking of search results by analyzing document relevance, and help in annotating data for learning-to-rank tasks in various learning contexts. However, this promising integration comes with its challenges, which include addressing potential biases and ethical issues in training models, managing the computational and other costs of incorporating LLMs into search services, and continuously updating LLM training with the ever-changing web content. We discuss these challenges and chart out required research directions to address them. We also discuss broader implications for service computing, such as scalability, privacy concerns, and the need to adapt search engine architectures for these advanced models.

\end{abstract}

\begin{IEEEkeywords}
Large Language Models (LLMs), Search Engines, Learning-to-Rank (LTR), and Retrieve-Augmented Generation (RAG)
\end{IEEEkeywords}
\vspace{-4mm}
}

\maketitle

\IEEEdisplaynontitleabstractindextext

\IEEEpeerreviewmaketitle

\ifCLASSOPTIONcompsoc

\section{Introduction}\label{sec:introduction}
The dawn of the Internet services age has brought forth a deluge of information, making the role of search engines more critical than ever in navigating this vast digital landscape~\cite{li1998toward,brin1998anatomy}. For instance, as of January 2024, the total number of websites worldwide has reached an impressive milestone of 1.079 billion. This figure marks a significant increase from the 185 million websites recorded 15 years ago, showcasing the exponential growth and expansion of the digital landscape over this period\footnote{https://siteefy.com/how-many-websites-are-there/}. However, as the complexity of user queries and the expectation for precise, contextually relevant, and up-to-date responses grow, traditional search technologies face mounting challenges in meeting these demands. Considerable advancements have been made in the fields of natural language processing (NLP) and information retrieval (IR) technologies~\cite{croft2010search,gao2015deep}. These efforts aim to enhance the ability of machines to accurately fetch content from the vast expanse of websites available online, efficiently store and index this content, comprehend user queries with higher precision, and deliver relevant, accurate, and current contents crawled from massive online websites, in an organized manner~\cite{hawking2001measuring,he2017scale}. 

On the other hand, Large Language Models (LLMs)-- the cornerstones of generative artificial intelligence (GenAI) have shown remarkable capabilities in understanding, generating, and augmenting human language~\cite{bubeck2023sparks,liu2023gpt}. The potential integration of LLMs with search engine services presents an exciting frontier in services computing, promising to significantly enhance search functionalities and redefine user interaction with digital information systems. For example, new Bing utilizes ChatGPT to perform Retrieval-Augmented Generation (RAG)~\cite{lewis2020retrieval} by injecting search results into the contexts of the LLM, generating comprehensive responses based on the most relevant and current information searched from its database\footnote{https://www.microsoft.com/en-us/edge/features/the-new-bing}. From the perspective of LLMs, this integration significantly enhances their accuracy and informativeness by allowing them to access and incorporate real-time data and diverse content from the web, thereby expanding their knowledge base beyond pre-training/fine-tuning datasets and enabling them to provide more accurate, contextually relevant, and up-to-date responses to user queries. Especially, search engines could help LLMs counter the hallucinations -- an innate of almost every LLM~\cite{shuster2021retrieval,foulds2024ragged}. From the perspectives of search engines, leveraging LLMs equipped with RAG capabilities enriches the user experience by offering more meticulous and contextually aware responses. This not only improves the search accuracy but also elevates the overall user experience in handling complex queries, thereby increasing user satisfaction and engagement with the platform~\cite{lewis2020retrieval,foulds2024ragged}.

In this work, we aim to explore the symbiotic relationship between LLMs and search engines, investigating how each can leverage the strengths of the other to overcome their respective limitations and to enhance thrier capabilities. As shown in Fig~\ref{fig:co-evolution}, the technologies driving the development of search engines and AI models have historically co-evolved. Both revolutionary streams of technology emerged around the same time, starting with the conceptualization of the Memex by Vannevar Bush and the pioneering work on artificial neurons by McCulloch and Pitts~\cite{nyce1991memex,mcculloch1943logical}. These foundational technologies paved the way for significant advancements, such as the World Wide Web (WWW) and PageRank for search engines, and Backpropagation, Recurrent Neural Networks (RNN), Convolutional Neural Networks (CNN), and Long Short-Term Memory (LSTM) models in AI from the 1980s to the 1990s~\cite{berners1994world,brin1998anatomy,lecun2015deep}. Following the historic win of AlexNet at the ImageNet competition in 2013~\cite{lecun2015deep}, Google elevated its retrieval and ranking components by integrating neural networks\footnote{https://blog.google/products/search/how-ai-powers-great-search-results/}. This propelled the continuous advancement of AI with the introduction of Transformer models with self-attention mechanisms, and BERT for enhanced query and content understanding in the late 2010s~\cite{vaswani2017attention,devlin2018bert}. More recently, OpenAI introduced the Generative Pre-trained Transformer (GPT) to start GenAI, and launched its groundbreaking online chatbot service, known as ChatGPT~\cite{openai2023gpt4}. Subsequently, Microsoft integrated ChatGPT into their search engine, forming the new Bing, which offers an advanced chat-based search experience in 2023.

\begin{figure}
    \centering
    \includegraphics[width=0.45\textwidth]{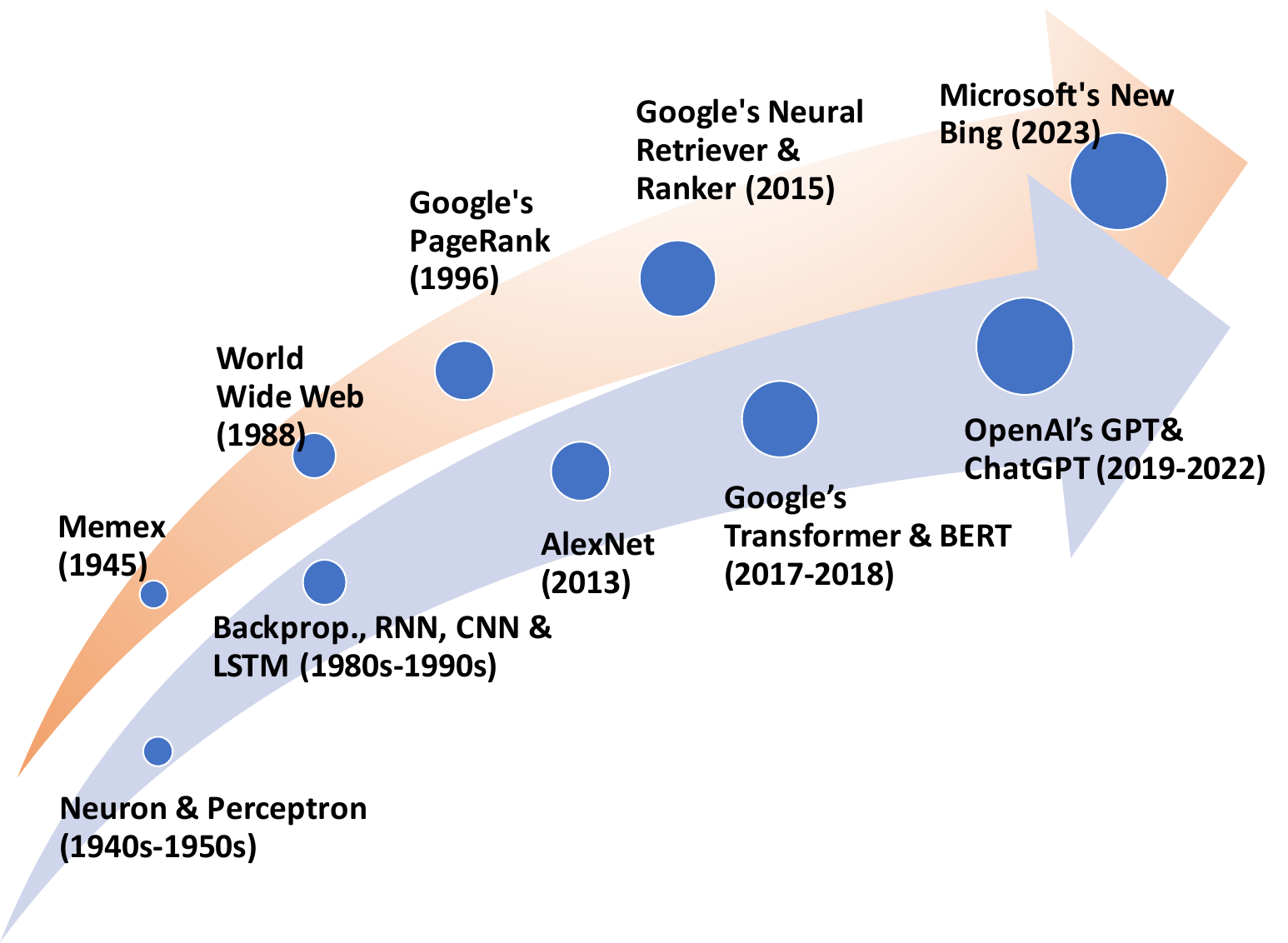}\vspace{-3mm}
    \caption{Technological Evolution of AI Models and Search Engine Technologies: Some of the key milestones achieved by AI and search engine (information retrieval) technologies.}
    \label{fig:co-evolution}
    \vspace{-5mm}
\end{figure}

In the context of services computing, the integration of LLMs and search engines is not merely an augmentation of existing capabilities but a paradigm shift towards creating more intelligent, efficient, and user-centric search services. The exploration is divided into two main themes: the benefits of enriching LLMs with search engine data and functionalities (\emph{Search4LLM}) and the enhancement of search performance through the capabilities of LLMs (\emph{LLM4Search}).
\begin{itemize}
    \item \textbf{\em Search4LLM}: Under this theme, we examine the process of leveraging the vast, diverse data repositories of search engines for the  pre-training and progressive fine-tuning of LLMs~\cite{yang2019xlnet}. This includes an examination of how high-quality, ranked documents can serve as an excellent source for training data, assisting LLMs in developing a better understanding of query contexts and improving their accuracy in generating relevant responses. Additionally, we focus on the potential of learning-to-rank (LTR) algorithms~\cite{liu2009learning} in refining  capabilities of LLMs to understand and prioritize information relevance. 
    
    \item  \textbf{\em LLM4Search}: Conversely, this part highlights the  impact that LLMs can have on search engine operations. This encompasses the utilization of LLMs for more effective content summarization~\cite{goyal2022news}, aiding in the indexing process~\cite{liu2022llamaindex}, and providing fine-grained query optimization techniques for superior search outcomes~\cite{ye2023enhancing}. Moreover, the potential of LLMs in analyzing document relevance for ranking purposes and facilitating data annotation in various LTR frameworks~\cite{wang2024simple,li2023s2phere,li2023coltr} is explored.
\end{itemize}

While this work examines a promising integration of LLMs with search engine services, it is beset by numerous challenges. These include the technical demands of deploying advanced models, ethical concerns, biases in model training, and the need for continuous updates to training datasets due to the evolving nature of web content. This study aims to offer groundbreaking insights and a systematic framework for future research and development in merging LLMs with search engines through a thorough investigation. Our exploration endeavors to augment the field of services computing, striving to develop smarter, more adaptive, and user-centric search services capable of adeptly managing the complexities of today's digital information landscape and offering the  user superior search experience. The key technical contributions of this research are summarized as follows: 
\begin{itemize}
    \item \textbf{\em Exploration of Innovative Utilization of Search Engine Data}: Investigates the potential of using broad and diverse datasets from search engines for the initial pre-training and subsequent fine-tuning of LLMs, enhancing their comprehension of queries and improving their accuracy in generating responses.
    
    \item \textbf{\em Exploring Leveraging High-Quality Ranked Documents in Training}: Examine the use of high-quality, ranked documents as superior sources of training data for LLMs, with the goal of improving their capability to deliver relevant and precise responses to user queries.
    
    \item \textbf{\em Advancement in LTR Technologies}: Investigates the application of LTR algorithms to augment the effectiveness of LLMs in assessing and prioritizing the relevance of information, thereby enhancing the precision of search results and response generation.
\end{itemize}
These contributions collectively represent significant advancements in both \emph{Search4LLM} and \emph{LLM4Search} themes.

\section{Backgrounds and Preliminaries}
In this section, we present the fundamentals of search engine services and LLMs to lay the groundwork for our research. 

\subsection{Search Engine Services}
In this section, we provide a concise review on search engine services. Referencing Figure~\ref{fig:search-service}, our analysis specifically concentrates on the architectural configuration of systems, the strategic implementation of algorithms, and the administration of evaluative experiments within a search engine.

\begin{figure}
    \centering
    \includegraphics[width=0.499\textwidth]{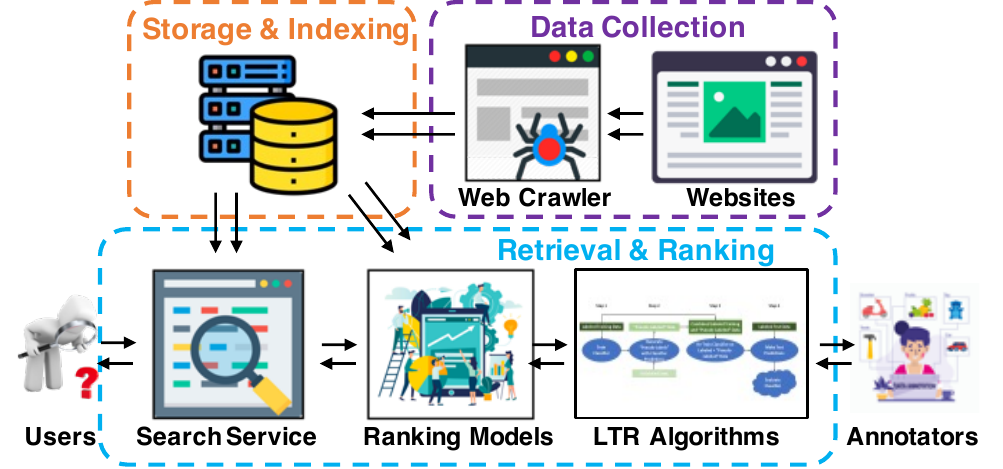}
    \caption{Architectural Design, Essential Components with Functionalities of a Common Search Engine Service}
    \label{fig:search-service}\vspace{-5mm}
\end{figure}

\subsubsection{Data Collection}
The performance of search engine services heavily relies on the gathering and examination of expansive online content. For this process, the use of efficient \textbf{web crawlers} is paramount. They systematically browse the World Wide Web to gather a wide variety of web resources including web pages, images, videos, and other multimedia content, which are crucial for maintaining the search engine's ability to provide comprehensive responses to user queries~\cite{kumar2017survey}.

The data collection process is complemented by \textbf{term extraction} module -- extracting key terms and phrases from the content. Term extraction leverages advanced text analysis and NLP techniques that identify and categorize important information, thereby refining the search engine's match between user queries and relevant documents. The optimization of this process is further enhanced by utilizing metadata like titles, descriptions, and tags, along with implementing sophisticated algorithms for entity recognition, semantic and sentiment analysis~\cite{skluzacek2018skluma,kumar2017survey}.

\subsubsection{Storage and Indexing}
Document storage and indexing form the backbone of a search engine's ability to quickly and accurately match and deliver search results. A critical part of this indexing process is the creation of an \textbf{inverted index}, a fundamental data structure that associates terms with the documents they are found in. This significantly reduces search time by narrowing down the search to documents containing the queried keywords~\cite{patil2011inverted}.

Additionally, \textbf{term weighting} strategies such as TF-IDF are implemented to rank terms within documents based on their frequency and relevance, improving the accuracy and relevance of search results by prioritizing highly informative terms~\cite{raghavan2008scoring}. These techniques ensure a precise match between user queries and indexed materials, significantly enhancing the search experience.

\subsubsection{Retrieval and Ranking}
Efficient document retrieval and ranking are important for delivering relevant and valuable search engine results. The primary stages in this process include:
\begin{itemize}
    \item \textbf{Query Processing}: The first step involves analyzing and potentially reformulating the user's query using advanced NLP techniques. This phase is crucial for understanding search intent and improving document retrieval effectiveness~\cite{tonellotto2018efficient}.
    
    \item \textbf{Relevance Scoring}: Each document is assessed and given a relevance score, based on criteria like query term frequency, document structure, and semantic content. This step quantifies the document's relevance to the query~\cite{yin2016ranking}.
    
    \item \textbf{Document Ranking}: Utilizing relevance scores and other factors (e.g., user metrics, site authority), algorithms like PageRank and machine learning models determine the document order, prioritizing the most pertinent results~\cite{liu2009learning,valizadegan2009learning}.
    
    \item \textbf{Search Results Personalization}: Personalization of results based on users' profile, search history, locations, and devices aims to enhance user satisfaction by tailoring outcomes adaptively~\cite{speretta2005personalized}.
    
    \item \textbf{Continuous Optimization}: The process is dynamically refined through A/B testing, user feedback, and technological progress to align with user preferences and content changes~\cite{mckeeth2011method}.
\end{itemize}

The ranking algorithms, particularly those based on Learning-to-Rank (LTR) models, are fundamental for search engines to sequence results with precision. LTR models, informed by user interactions and feedback, employ different approaches to ranking:
\begin{itemize}
    \item \textbf{Pointwise approaches}: View ranking as a regression or classification to predict individual document scores.
    
    \item \textbf{Pairwise approaches}: Focus on the relative ranking between document pairs~\cite{cao2007learning}.
    
    \item \textbf{Listwise approaches}: Aim to optimize the entire result list's order~\cite{xia2008listwise}.
\end{itemize}
Listwise methods are notably effective in achieving user satisfaction~\cite{jiang2016correlation}.

The development of LTR models is influenced by the availability of human-annotated data, leading to various methodologies:
\begin{itemize}
    \item \textbf{Active LTR} prioritizes annotating query-document pairs with uncertain predictions for efficient model training with fewer examples~\cite{wang2024simple}.
    
    \item \textbf{Semi-Supervised LTR} combines limited labeled data with larger unlabeled datasets to enhance model training, employing strategies like self-training for effective use of annotations~\cite{li2023coltr,li2023s2phere}.
    
    \item \textbf{Pretrain-Finetuned LTR} involves pre-training on vast datasets followed by fine-tuning with annotated query-document pairs. This approach significantly improves ranking accuracy and data usage efficiency~\cite{li2023towards,li2024gs2p}.

\end{itemize}
Selecting among these methods is dictated by the specific LTR challenges and objectives within search engines~\cite{werner2022review}.  

\subsubsection{Evaluation for Search Engine Services}
We outline a technical framework for assessing search engine performance, incorporating critical methodologies and metrics for comprehensive experimentation and analysis. 

A/B testing, which is crucial for ongoing enhancement in search engine performance, involves comparing two variations of a search engine to determine the one that performs better~\cite{quin2024b}. The A/B testing protocol involves:
\begin{itemize}
    \item \textbf{Establishing Objectives}: Defining measurable goals, such as boosting click-through rates or search result relevance~\cite{lyon2015b}.
    
    \item \textbf{Creating Variants}: Developing a control version (A) versus an experimental version (B), with the latter introducing a new ranking model or feature~\cite{kohavi2015online}.
    
    \item \textbf{Segmenting Users}: Randomly allocating users to either variant to ensure statistical comparability and isolate the effects of changes~\cite{lyon2015b}.
    
    \item \textbf{Test Implementation}: Running the test until reaching statistical significance, while collecting performance data for both variants~\cite{chen2019b}.
\end{itemize}

Evaluating search engine performance necessitates metrics that accurately reflect user satisfaction with search results. Essential KPIs include:
\begin{itemize}
    \item \textbf{Precision at Top-k Results (P@k)} and \textbf{Normalized Discounted Cumulative Gain (NDCG)} for ranking accuracy of the search outcomes~\cite{sujatha2011precision,wang2013theoretical}.
    
    \item \textbf{Mean Reciprocal Rank (MRR)} for the speed of relevant content retrieval~\cite{chowdhury2002automatic}.
    
    \item \textbf{Click-through Rate (CTR)} and \textbf{User Satisfaction} for gauging engagement and satisfaction~\cite{dan2016measuring}.
    
    \item \textbf{Conversion Metrics} for assessing the economic impact of changes on commercial search engines~\cite{ghose2008empirical}.
\end{itemize}

Through A/B testing and rigorous evaluation with these KPIs, search engine developers can make informed decisions to enhance user experience, relevance, and achieve business objectives, ensuring ongoing improvement and optimization of search technology.

\begin{figure}
    \centering
    \includegraphics[width=0.5\textwidth]{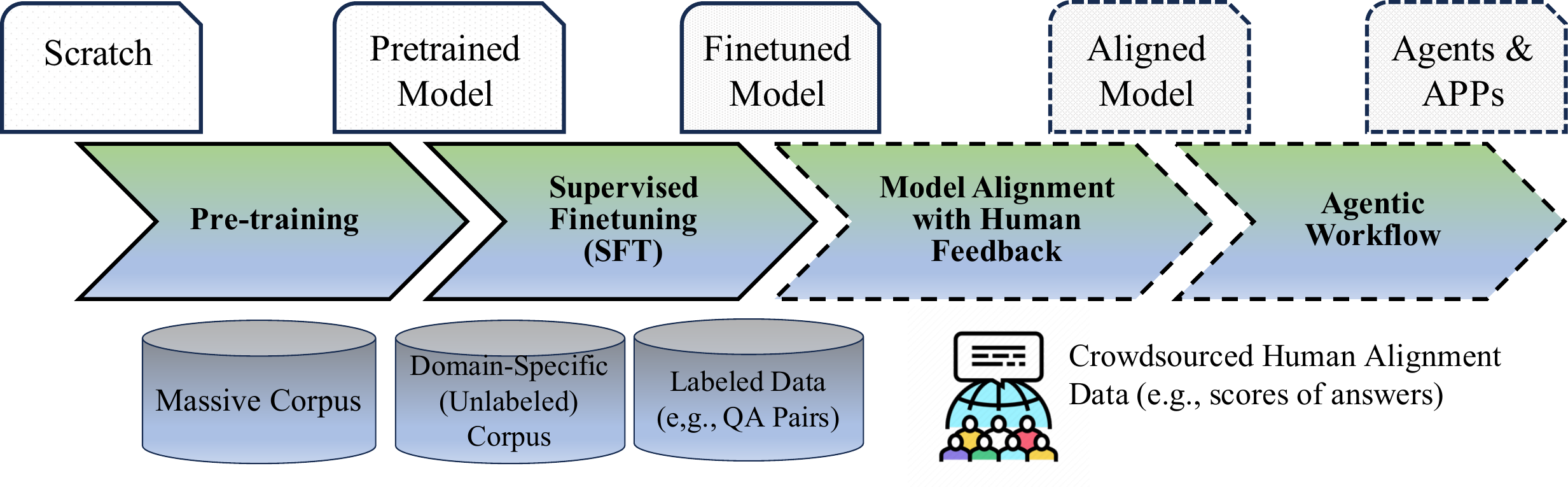}
    \caption{The Life-cycle of LLMs: Pre-training, supervised fine-tuning, model algiments with human feedback, and building applications with agents.}\vspace{-5mm}
    \label{fig:model-lifecycle}
\end{figure}

\subsection{Large Language Models (LLMs)}
Large Language Models (LLMs) represent a significant advancement in the field of natural language processing (NLP) and artificial intelligence (AI). These models have fundamentally altered the landscape of computational linguistics, enabling a wide array of applications that range from text generation to complex question-answering systems~\cite{zhao2023survey}. This section delves into the foundational architectures of LLMs and the full life-cycle (shown in Fig.~\ref{fig:model-lifecycle}), ranging from pre-training, to supervised fine-tuning, to model alignments, and to agent-based applications, which elevate the capabilities of LLMs. 


\subsubsection{Foundation Models of LLMs}
Introduced by ``Attention is All You Need''~\cite{vaswani2017attention}, transformers revolutionized NLP by utilizing self-attention mechanisms over recurrent layers. This innovation allows simultaneous word processing in sentences, enhancing efficiency and linguistic comprehension.
We, here, compare encoder-only, decoder-only, and encoder-decoder models of transformers, exemplified by BERT, GPT, and BART, respectively~\cite{ghojogh2020attention,lin2022survey}.
\begin{itemize}
    \item \textbf{Encoder-Only Models}: BERT exemplifies this category with its bidirectional training enhancing context understanding. Its encoder transforms input sequences into contextualized representations, aiding in various NLP tasks~\cite{kenton2019bert}.
    \item \textbf{Decoder-Only Models}: GPT and related models emphasize text generation through stacked decoder layers. They predict subsequent words based on previous ones, enabling coherent text generation~\cite{achiam2023gpt}.    
    \item \textbf{Encoder-Decoder Models}: BART combines both approaches for robust language understanding and generation. This architecture supports a wide range of tasks including summarization and translation~\cite{lewis2020bart}.
\end{itemize}
Each model type, from encoder-only to encoder-decoder, offers unique capabilities for specific NLP applications. The evolution from basic transformer models to specialized ones like BERT, GPT, and BART highlights rapid advancements in NLP technology~\cite{ghojogh2020attention,lin2022survey}.

\subsubsection{Pre-training Models}
Pre-training models undergo essential tasks to understand and generate natural language effectively. A condensed overview of these tasks is as follows:
\begin{itemize}
    \item \textbf{Masked Language Modeling (MLM)}: This method conceals specific words in a text, provoking the model to predict the omitted words based on the context. This process is critical for comprehending the bidirectional context~\cite{kenton2019bert}.

    \item \textbf{Next Token Prediction}:  It involves predicting subsequent words in a sequence, teaching the model the likelihood of word sequences. This is essential for models aimed at text generation like GPT~\cite{bengio2000neural,liu2023gpt}.  

    \item \textbf{Next Sentence Prediction (NSP)}: With this task, models are trained to assess whether a sentence logically follows another, enhancing sentence-level comprehension for tasks like text classification~\cite{sun2022nsp}.

    \item \textbf{Permutation Language Modeling (PLM)}: This task is unique to models like XLNet where word order is scrambled for the model to predict the original arrangement, aiding in non-linear understanding of contexts~\cite{yang2019xlnet}.

    \item \textbf{Sentence Order Prediction (SOP)}: An advancement of NSP, where models reorder shuffled sentences in a text, improving their grasp on narrative flow and long-range dependencies~\cite{lan2019albert}.

    \item \textbf{Contrastive Learning}: This task focuses on differentiating between correct and corrupted input versions, refining the models' semantic comprehension~\cite{clark2019electra}.
\end{itemize}
These tasks collectively prepare Large Language Models (LLMs) for a wide array of NLP applications by fostering a robust linguistic foundation.

\subsubsection{Supervised Fine-tuning (SFT) and Alignments}
Following the broad-based learning in the pre-training stage, LLMs undergo Supervised Fine-tuning (SFT) to enhance their capabilities for particular applications. Later, model alignments, such as Reinforcement Learning from Human Feedback (RLHF), would be carried out to adjust the model's outputs to closely match human expectations and norms, thereby improving the model's efficacy, accuracy, and even ethical considerations in its applications~\cite{bai2022training}.

SFT is a powerful technique for optimizing LLMs to perform specific tasks with enhanced accuracy and performance. This process involves leveraging the pre-existing knowledge of the LLM, gained through pre-training on extensive datasets, and adapting it to excel in targeted applications. 
\begin{itemize}
    \item \textbf{Data Preparation}: The process begins by selecting task-specific, labeled datasets that align with the intended application of the LLM. This data could range from specialized corpora in sectors like healthcare or finance to structured question-answer pairs for tasks such as question-answering (QA)~\cite{sun2022nsp}.

    \item \textbf{Training Procedure}: SFT capitalizes token sequence prediction tasks (e.g., question-answering), enhancing the model's adaptability and accuracy without sacrificing generalizability~\cite{sun2022nsp}.

\end{itemize}

RLHF involves several steps designed to align the model's outputs with qualitative judgments or desired behaviors as determined by human feedback~\cite{bai2022training,ouyang2022training}. Key components include:
\begin{enumerate}
    \item \textbf{Reward Modeling}: Training a model to predict preferred outcomes by evaluating model outputs against human judgments, aligning predictions with human values~\cite{moskovitz2023confronting}.

    \item \textbf{Proximal Policy Optimization (PPO)}: Employing PPO to update decision-making policies towards maximum reward outcomes, ensuring effective learning from complex feedback~\cite{wu2023pairwise}.

    \item \textbf{Fine-tuning with Human Feedback}: Continual fine-tuning using human feedback on new samples to refine both the LLM and the reward model, enhancing model alignment with human expectations~\cite{bai2022training}.
\end{enumerate}
By integrating model SFT and alignments, LLMs achieve superior performance, ethical soundness, and practical value in applications~\cite{dai2023safe,casper2023open}.

\subsubsection{LLM Extensions and Usages}
In the era of foundation models, LLMs have emerged as versatile tools with impactful applications across different domains. Harnessing the power of LLMs, notable advancements have been witnessed in the domains of Prompts, Reasoners, and Agents. Let's delve into each of these perspectives to explore the diverse applications of LLMs.
\begin{itemize}
    \item \textbf{Prompts}: Prompting techniques are essential for effectively utilizing LLMs, enabling them to comprehend and react to user needs. Prompt engineering allows for the customization of LLM output through advanced techniques such as few-shot and zero-shot (in-context) learning, thus improving task adaptability~\cite{xie2021explanation,dong2022survey}. Additionally, prompts open up possibilities for innovative and dynamic interactions with LLMs, enhancing user engagement~\cite{dang2022prompt,meincke2024prompting}.
    
    \item \textbf{Reasoners}: LLMs, powered by reasoning techniques like chain-of-thought (CoT) and tree-of-thought (ToT), excel in complex problem-solving by mimicking human reasoning processes~\cite{wei2022chain,yao2024tree}. These methods enable LLMs to extend their knowledge base, stay current with information, and address bias and fairness in their responses~\cite{ling2023knowledge,chisca2024prompting,ma2024fairness}.
    
    \item \textbf{Agents}: Acting as autonomous agents, LLMs autonomously perform tasks, interact with external tools, and learn to improve their performance over time with minimal human intervention~\cite{albrecht2018autonomous,li2023autonomous}. These agents are notable for their memory and planning abilities, collaboration potential, and the capacity for customization, making them versatile across various applications~\cite{li2023autonomous,wang2024survey,hong2023metagpt,yang2023auto,xiong2023natural,shen2024hugginggpt}.
\end{itemize}
In essence, the applications of LLMs through prompts, reasoning frameworks, and autonomous agents showcases their broad capabilities and potential for innovation across different domains. Continual advancements in this sphere promise to further enhance LLM utility and versatility.

\section{Search4LLM: Enhancing LLMs with Search Engine Services}
In this section, we present our vision under the theme of \emph{Search4LLM}, where we specifically examine how search engine services can significantly enhance the full \emph{life-cycle of LLMs} from pre-training, to fine-tuning and model alignments, and to applications of LLMs. An overview of this theme has been illustrated in Fig.~\ref{fig:search4llm}.

\begin{figure*}
    \centering
    \includegraphics[width=0.75\textwidth]{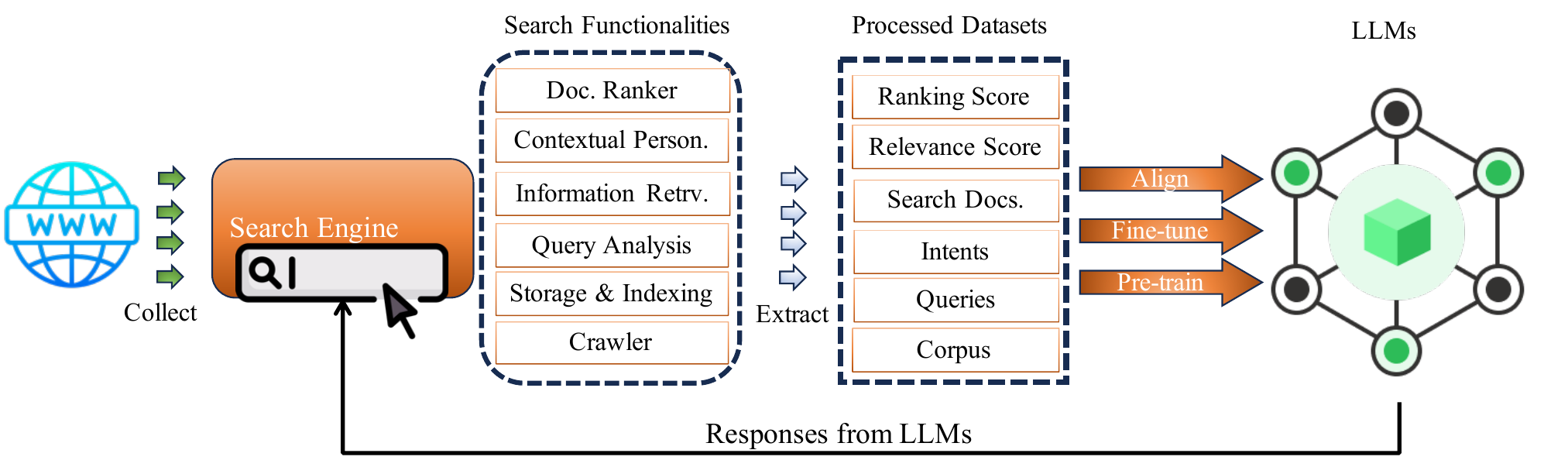}\vspace{-3mm}
    \caption{An Overview of \emph{Search4LLM} Theme: leveraging the search engine functionalities to process the data crawled from web and responses from LLMs, providing datasets for pre-training, supervised fine-tuning, and model alignments.}
    \label{fig:search4llm}\vspace{-5mm}
\end{figure*}

\subsection{Enhanced LLM Pre-training}
Search engines play a critical role in the pre-training phase of LLMs. This initial phase is foundational, setting the groundwork upon which further model-specific training is built. The utility of search engines in this context cannot be overstated, as they provide a unique and powerful means of collecting, categorizing, and indexing vast swathes of online content. Such capabilities directly impact the quality and efficacy of LLM pre-training in several key ways.

\subsubsection{Collection of Massive Online Contents as Corpus}
At the core of LLM pre-training is the need for extensive and varied datasets. The functionality of search engines in scouring the internet enables the collection of a vast array of data from multiple sources, encapsulating a diverse range of languages, formats -- including HTML pages, PDFs, and text files -- and topics from scientific research papers to literary works and current news articles~\cite{kausar2013web,kumar2017survey}. 

The wide spectrum of content, collected and aggregated by search engines, serves as an ideal corpus for LLM pre-training. It allows the resulting language models to develop a comprehensive understanding of language patterns, semantics, and syntax. Such comprehensive corpora is instrumental in spanning the vast landscape of human language and applications, thereby ensuring the model's broad applicability across different areas. This strategic compilation not only fosters a deeper comprehension of language intricacies but also solidifies the foundation for creating models that can adeptly navigate the complexities and subtleties of human expression found across the wide-ranging contents at web-scale~\cite{penedo2024refinedweb}.

\subsubsection{Indexing Corpus by Domains and Quality of Texts}


During the pre-training phase of Language Learning Models (LLMs), the act of categorizing the corpus according to various domains and conducting an evaluation of text quality serves a pivotal role in guaranteeing a balanced data distribution. This practice is integral to the development of a comprehensive and impartial model. This methodology involves a rigorous selection of the dataset encompassing a myriad of domains such as news, research reports, literature, and colloquial internet language, all while emphasizing the diversity, reliability, and authority of contents through specific quality indicators~\cite{lee2020beyond}. 

In this way, LLMs not only benefit from a diversified learning experience, minimizing the risk of domain biases and the over-representation of certain linguistic styles, but also are able to understand and generate language with a remarkable balance and broad applicability~\cite{liu2022improved}. This method does not merely serve as a strategic preference but emerges as an essential strategy in the development of comprehensive, equitable LLMs capable of navigating through the extensive yet unique landscape of human communication~\cite{bhardwaj2023pre}.

\subsubsection{Supporting Continuous Model Improvement}
Search engines operate as dynamic repositories of information, continuously updated by web crawlers that traverse the internet to index new and revised content. This ever-evolving corpus of information serves as a vital resource for the continuous improvement of LLMs, particularly in keeping these models relevant, accurate, and reflective of current language usage and trends~\cite{lewandowski2006freshness}. 

Take a LLM-backed chatbot as an example. As global events unfold or new discoveries are made, the chatbot must understand and provide information on these topical events. By regularly updating the LLM with content indexed by search engines—such as news articles and reports on these recent events--the model remains competent in delivering timely and relevant responses to user inquiries~\cite{vu2023freshllms}.

\subsection{Enhanced LLM Fine-tuning}
Search engines play a key role in the fine-tuning process of LLMs, enhancing their ability to interact with users and provide accurate, contextually relevant responses upon specific domains. This process leverages the advanced capabilities of search engines, including query rewriting, the analysis of user interactions, and the utilization of domain-specific content. By integrating these elements into the fine-tuning of LLMs, the models can significantly improve in three aspects as follows.

\subsubsection{Learning to Follow User Instructions}
One of the primary enhancements search engines offer to LLM fine-tuning involves teaching the model to recognize and interpret users' intentions. The capability of instruction-following could be achieved through the mechanism of query rewriting, where search algorithms adjust or reformulate user queries to better capture the user's intent~\cite{tucker2010methods}. 

By analyzing patterns in query rewriting, LLMs can learn to infer the underlying intentions behind users' queries, enabling them to respond more accurately and helpfully. This technique not only improves the model's comprehension of user requests but also its ability to engage in more intuitive and efficient dialogue~\cite{ouyang2022training}.

\subsubsection{Learning to Answer Questions}
The fine-tuning process also capitalizes on structuring datasets that simulate the question-answering dynamic, utilizing actual search queries as the basis for generating questions and selecting top-relevant content or user-most-clicked items as the corresponding answers. By doing so, the model is trained on real-world examples of how users phrase queries and what information they find most useful, based on search engine results and click-through data. This approach provides the LLM with a rich dataset reflective of genuine user interactions, enabling the model to better understand and structure its responses in a manner that aligns with user expectations and the typical flow of information retrieval~\cite{puri2020training,kim2023sure}.
Fig.~\ref{fig:qa-sft} illustrates an example of leveraging search queries and top search results, collected from a search engine, to synthesize question-answer (QA) pairs for SFT. Classic NLP techniques or prompt-based tuning with LLMs could be used to organize search results into an answer to the question in the query.

\begin{figure}
    \centering
    \includegraphics[width=0.5\textwidth]{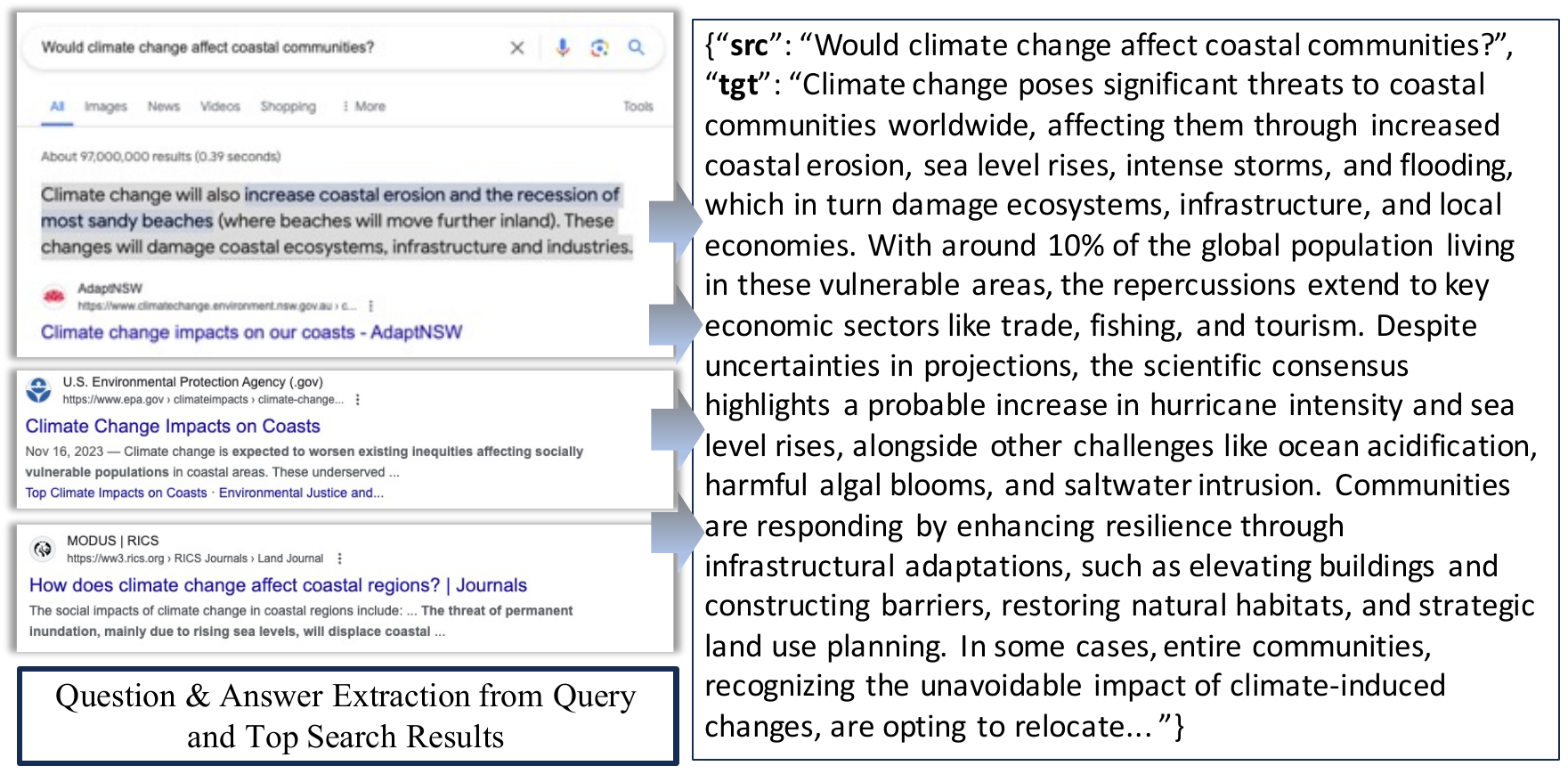}
    \caption{Extracting Questions and Answers for SFT from Search Queries and Top Search Results}
    \label{fig:qa-sft}\vspace{-5mm}
\end{figure}

\subsubsection{Incorporating Domain-specific Knowledge}
Finally, the utilization of domain-specific queries and content curated by search engines serves as a cornerstone for fine-tuning LLMs with specialized knowledge. This involves leveraging search engine capabilities to gather and categorize information specific to distinct fields or industries, such as medicine, law, or technology~\cite{pham2018learning}. 

By fine-tuning the model on datasets comprised of domain-relevant queries and authoritative content, the LLM acquires an in-depth understanding of sector-specific terminologies, concepts, and commonly sought information. This process not only enhances the LLM's expertise in various domains but also its ability to deliver precise, expert-level answers to queries within those specific areas~\cite{huang2023dsqa,ge2024openagi}.

\subsection{Enhanced LLM Alignment}
Search engines, with their web crawling technologies and advanced semantic algorithms, offer invaluable tools for enhancing the alignment of LLMs with human values and improving the relevance and quality of their outputs. These technologies, developed and refined through long-term operations, provide a framework for ensuring that LLMs can represent human values accurately, prioritize content relevance, and maintain high-quality output. Fig.~\ref{fig:search-align} illustrates the framework of leveraging these components to provide feedbacks for model alignment. Through integrating specific functionalities with search engines, LLMs can achieve a greater degree of alignment as follows.

\subsubsection{Semantic Relevance Alignment}
The LTR system, an integral component of search engines, is engineered to organize and display search results according to their semantic significance in relation to the user's search query~\cite{liu2009learning,werner2022review}. This functionality can be particularly beneficial when LLMs produce multiple outputs in response to a single input, a common occurrence with the application of expert or decoding methods aimed at enhancing response diversity. 

By applying the LTR system to these sets of results, it is possible to rank the outputs in order of their relevance and utility regarding the initial query. This practice ensures that the most relevant information is prioritized, helping users to access the most accurate and helpful content more efficiently~\cite{dupret2010model}.

\subsubsection{Content Value Alignment}
Search engines deploy elaborate crawling algorithms capable of identifying content that may be harmful, such as hatred, pornography, or violence, even when the content doesn't clearly seem sensitive or offensive at first glance. This capability stems from long-term exposure to vast quantities of online media and the continuous refinement of content evaluation models~\cite{warner2012detecting}. 

Integrating above modules or functionalities (already existing in search engines) into the LLM fine-tuning process allows for the incorporation of human values at the core of model alignment. By leveraging the search engine's ability to discern and filter out undesirable content, LLMs can be trained or corrected to avoid generating or promoting material that contradicts widely accepted human values, thus ensuring the model's outputs are aligned with ethical and societal norms~\cite{liu2023trustworthy,zhang2024heterogeneous}.

\subsubsection{Content Quality Alignment}
Search engines are commonly equipped with models that evaluate the quality of online content, often trained using extensive datasets of users' click-through data. These models assess various aspects of content, such as its credibility, informativeness, and user engagement for quality-based search or ranking~\cite{mandl2006implementation,schultheiss2022does}. 

By applying above evaluation models to review and rate the content generated by LLMs, search engines can provide critical feedback for the continuous alignment and improvement of the models. This feedback loop enables the identification of content quality issues, guiding subsequent fine-tuning efforts to enhance the overall quality of LLM outputs. In turn, this process contributes to the optimization of LLMs, ensuring they produce high-quality, relevant information that meets user expectations~\cite{liu2023g,song2023preference}.

\begin{figure*}
    \centering
    \includegraphics[width=0.8\textwidth]{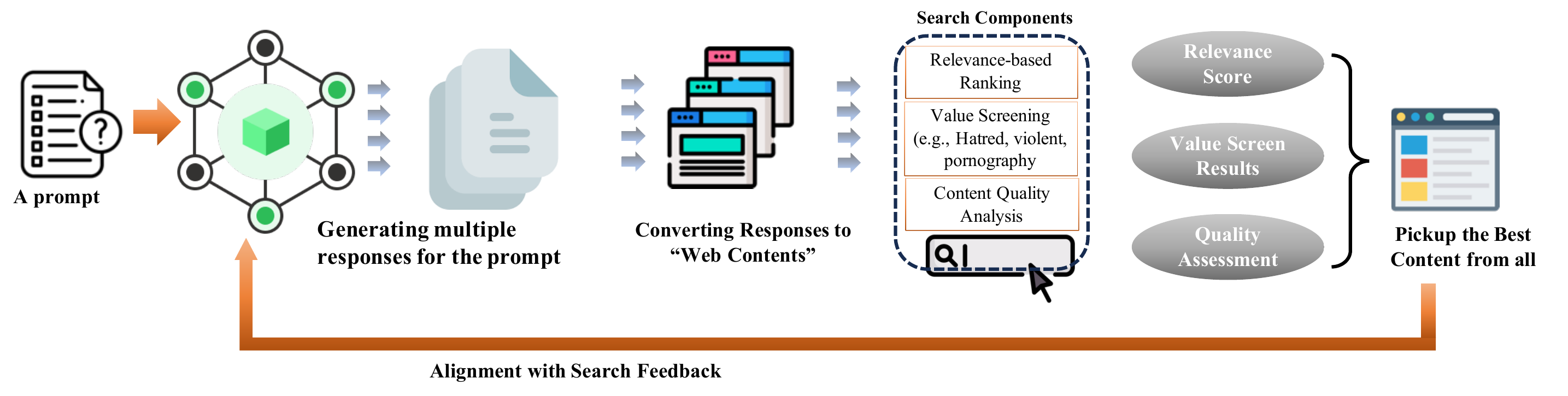}\vspace{-3mm}
    \caption{Using Relevance, Content Quality, and Value Screening Components to Align Models}\vspace{-5mm}
    \label{fig:search-align}
\end{figure*}
\subsection{Enhanced LLM Applications}
Incorporating search engine capabilities significantly enhances LLMs by addressing their key limitations from applications' perspectives, such as the lack of real-time information, difficulty with out-of-distribution questions, and constraints within specific domains. 

\subsubsection{Real-time Information Provision}
LLMs are constrained by the datasets they are trained on, which, due to the extensive time required for pre-training, fine-tuning, and subsequent updates, often lack current information. Search engines, on the other hand, offer a conduit to real-time data across various domains. By leveraging retrieval-augmented generation (RAG)~\cite{lewis2020retrieval}, LLMs can dynamically integrate search-engine-sourced information into their responses~\cite{kelly2023bing}. 

Specifically, RAG involves executing real-time searches based on the input query and fusing the retrieved information with the LLM's generated content, thus enabling the model to provide up-to-date answers and insights~\cite{foulds2024ragged}. For example, when asking GPT-4 the question ``Today's weather in Washington DC.'' without access to the Internet, GPT-4 would respond the user with notice of non-access to real-time data. However, when including the top one result of the same query from Google search into the context of prompting for RAG, GPT-4 would respond the user with accurate information. 

\subsubsection{Cross-domain Question Answering}
LLMs, whether general-purpose or fine-tuned for specific domains, may struggle with questions that lie outside their trained datasets or knowledge domains, known as out-of-distribution or out-of-domain queries~\cite{yuan2024revisiting,wang2024beyond}. 

In such scenarios, search engines can serve as a powerful tool to supplement the LLMs' responses by providing cross-domain information. Specifically, when an LLM encounters a query it is ill-equipped to answer due to domain limitations, it can utilize search engines to fetch pertinent information from a broader spectrum of knowledge. This not only expands the range of questions the LLM can handle but also enhances the depth and accuracy of its responses, making the model more versatile and capable of tackling a wider array of subjects~\cite{kelly2023bing}.

\subsubsection{Addressing Miscellaneous Limitations}
Beyond real-time updates and cross-domain supplementation, search engines can assist LLMs in various other aspects. For instance, improving the model's ability to discern user intent by analyzing search patterns and query refinements, bolstering content quality through insights derived from user engagement metrics, and even refining the model's ethical and factual alignment by filtering out unreliable sources. 

In addition, the broad and continuously updated dataset a search engine handles provides a wealth of supplementary information that can be used to train and enhance LLMs in effective ways, addressing a range of miscellaneous limitations that might not be readily apparent during the initial model development stages.

\subsection{Summary of Search4LLM Research}
The introduction of search engine functionalities into LLMs presents a revolutionary stride in the development of AI, particularly in automating the procedures of massive data collection and fine-grained data production. This synergy provides a robust framework for enhancing the models' capacity to interpret user intentions, generate relevant responses, and apply specialized knowledge across diverse domains. Below, we delve into several key points that highlight the significant achievements and future prospects of integrating search engine capabilities into LLMs:
\begin{itemize}
    \item \textbf{Enhanced Understanding of User Intentions}: The use of query rewriting techniques within LLMs enables a more profound comprehension of what users are actually seeking. This advancement allows for a deep understanding of queries, catering to the specific needs and contextual inquiries of the users.
    
    \item \textbf{Augmented Answer Structuring}: Leveraging real-world search data, LLMs can now structure answers in a more coherent and informative manner. This not only enhances the utility of responses provided to user queries but also ensures that the information is presented in an easily digestible format, making it more accessible to users.
    
    \item \textbf{Application of Domain-Specific Knowledge}: By incorporating domain-specific content and expertise into their frameworks, LLMs can offer precise and contextually relevant answers. This significantly elevates their proficiency in handling inquiries that require specialized knowledge or expertise.
    
    \item \textbf{Optimization of Model Alignment with Human Values}: The integration facilitates a comprehensive approach to aligning LLM outputs with ethical standards and human values. Through content value alignment, learning-to-rank systems for prioritizing outputs, and utilizing quality assessment models for feedback, LLMs can achieve a balance between accuracy, ethical considerations, and user satisfaction.
    
    \item \textbf{Relevance and Accuracy Adjustment}: The collaboration between search engines and LLMs introduces mechanisms like retrieval-augmented generation, which significantly boosts the models' accuracy and relevance. This is particularly vital in overcoming challenges related to real-time data provision, domain-specific knowledge application, and addressing out-of-distribution queries.
    
    \item \textbf{Versatility and Dynamic Responsiveness}: With the integration of search engine technologies, LLMs exhibit unprecedented versatility and adaptability. They become more adept at navigating the complex and constantly changing landscape of human knowledge and communication, effectively managing cross-domain inquiries and providing up-to-date information.
\end{itemize}
In summary, it is our unique vision to incorporate search engine functionalities in the full life-cycle of LLMs (pre-training, fine-tuning, alignment and applications). As we move forward, this synergy between search engine capabilities and LLMs heralds a new era in AI, characterized by models that are more dynamic, responsive, and comprehensive, embodying a significant leap towards achieving AGI.

\begin{figure*}
    \centering
    \includegraphics[width=0.85\textwidth]{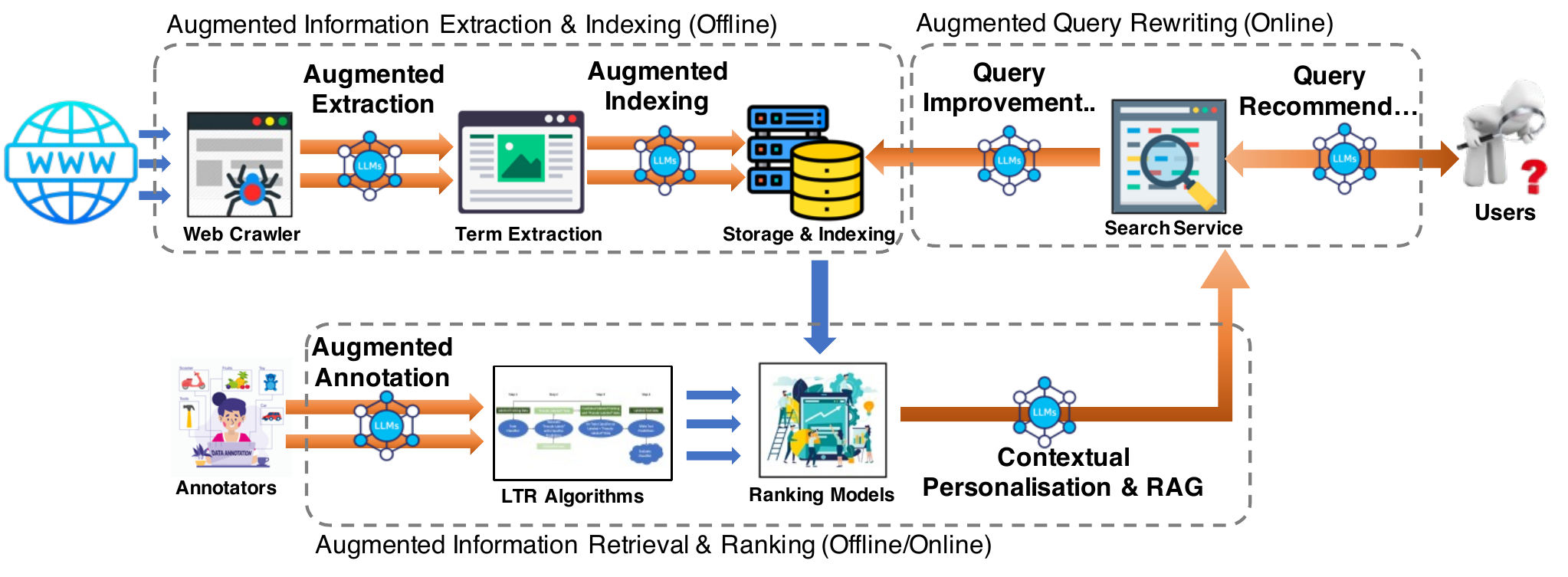}\vspace{-3mm}
    \caption{An Overview of \emph{LLM4Search} Theme: Leveraging LLMs to augment information extraction \& indexing, query rewriting \& improvement, and information retrieval \& ranking in online/offline manners.}
    \label{fig:llm4search}\vspace{-5mm}
\end{figure*}

\section{LLM4Search: Augmenting Search Engines with LLMs}
In this section, we present our vision under the theme of \emph{LLM4Search}, where we specifically examine how large language models (LLMs) can significantly augment LLMs in terms of query understanding, information extraction \& retrieval, and content ranking for web search. An overview of this theme has been illustrated in Fig.~\ref{fig:llm4search}.

\subsection{Augmented Query Rewriting}
The adoption of LLMs into search engine services has the potential to augment the rewriting process of search queries, thereby improving user experiences and search result relevance~\cite{ye2023enhancing,liu2024query,dhole2024genqrensemble}, in several ways as follows.

\subsubsection{Query Recommendation and Completion}
LLMs can significantly enhance query recommendation and completion functionalities in search engines by leveraging their deep understanding of language and context~\cite{anand2023query}. When a user begins typing a query, LLMs can analyze the partial input and generate highly relevant keyword suggestions and complete query predictions.
\begin{itemize}
    \item \textbf{Query Completion}: LLMs can comprehend the semantic meaning behind partial queries, allowing them to predict the user's intent and suggest relevant keywords or phrases that align with the intended search~\cite{desilva2024}. 
    \item \textbf{Query Recommendation}: LLMs can be fine-tuned on search query logs and user behavior data to identify trending keywords or popular patterns of queries. This information can be incorporated into the recommendation system, ensuring that suggested keywords and query completions reflect current user interests and preferences~\cite{desilva2024,mishra2023balancedqr}.
\end{itemize}

\subsubsection{Query Correction and Improvement}
Language Learning Models (LLMs) can serve a fundamental function in augmenting the capabilities of query correction within search engine systems. By leveraging their understanding of language and ability to identify and rectify errors, LLMs can assist users in refining their queries, even when faced with misspellings, grammatical errors, or incorrect inputs. Specifically, LLMs can be trained on large-scale text corpora to recognize and correct common spelling errors and grammatical errors in user queries. By understanding the structure and syntax of language, LLMs can suggest accurate spelling corrections or grammatically correct alternatives, ensuring that the search engine handle the query and retrieves relevant results~\cite{desilva2024,dhole2024genqrensemble}.

\subsubsection{Contextualized and Personalized Query Extension}
LLMs can significantly enhance the contextualization and personalization of query extensions in search engines. By leveraging information from cookies and browsing/search history, LLMs can tailor query extensions to individual users, providing a more relevant, personalized, and context-aware search experience~\cite{wang2023query2doc,li2023agent4ranking}.

Specifically, LLMs can analyze user-specific data, such as browsing history, search patterns, and preferences, to build comprehensive user profiles. These profiles can be used to understand the user's interests, expertise level, and search behavior, enabling personalized query extensions that align with their specific needs. Furthermore, LLMs can examine the context surrounding a user's query, including the current browsing session, previous searches, and the content of the web pages visited. By understanding the broader context, LLMs can extend queries that are highly relevant to the user's current information-seeking task~\cite{zhou2023unified,ye2023enhancing,liu2024query}.

\subsection{Augmented Information Extraction and Indexing}
LLMs stand at the forefront of transforming search engines' approach to information extraction and document indexing. LLMs, with their advanced understanding of natural language processing, can significantly improve the precision and relevance of the indexing process. 

\subsubsection{Terms Extraction and Summarization for Indexing}
LLMs possess the inherent capability to understand and interpret the contextual meaning and detailed information of text on web pages. This comprehension plays a vital role in pulling out an exact set of index terms and succinctly summarizing the content, both of which are key procedures in the task of document indexing~\cite{maoro2023leveraging}, as follows.
\begin{itemize}
    \item \textbf{Term Extraction}: By deploying LLMs, search engines can comprehend every webpage in depth, distinguishing crucial information from generic data. This discernment allows for the extraction of meaningful and precise index terms that accurately reflect the page's content~\cite{giguere2023leveraging,maragheh2023llm,goel2023llms}. 
    \item \textbf{Content Summary}: LLMs can generate succinct and informative summaries of web content. These summaries provide a quick overview of the webpage, aiding in the efficient categorization and retrieval of documents. This capability is particularly beneficial for users and search engines alike, offering a glimpse into the content without the need to parse through the entire document~\cite{xiao2023enhancing,zhang2023summit}.
\end{itemize}
Fig~\ref{fig:extract-term} illustrates an example of terms extraction and summarization for indexing purposes. With the original content provided in-context of the prompt, GPT-4 could respond the terms extracted and a snippet for summarization. Obviously, one could run the prompt with LLMs multiple times to diversity the extraction and summarization results.

\begin{figure}
    \centering
    \includegraphics[width=0.5\textwidth]{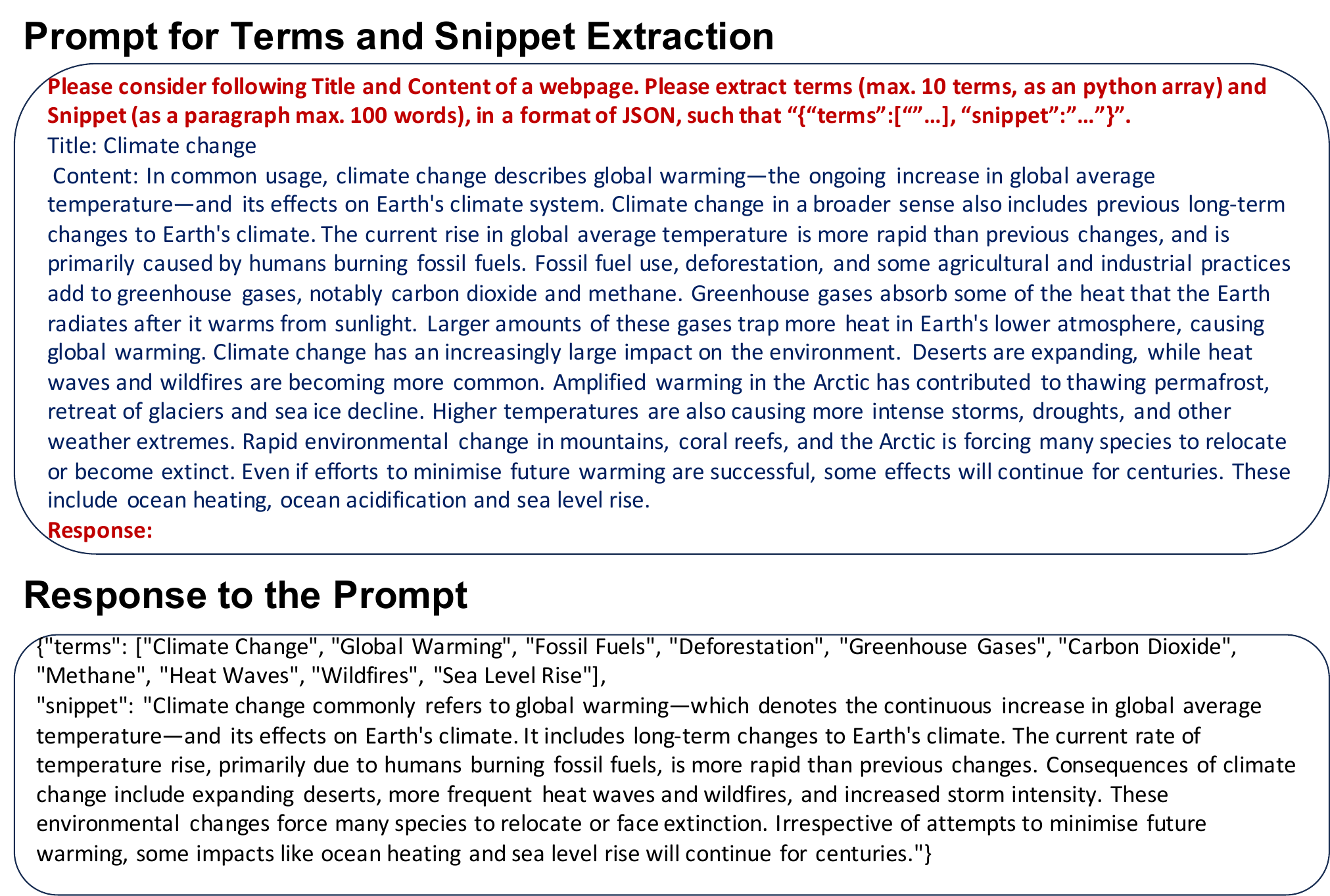}
    \caption{An example of Prompts and Responses for Terms Extraction and Summarization for Indexing Purposes}
    \label{fig:extract-term}\vspace{-5mm}
\end{figure}

\subsubsection{Semantic Labeling and Categorization for Indexing}
The capability of LLMs to measure the semantic distance or similarity between web pages is revolutionary, providing a reliable approach to automatically labeling and categorizing web pages based on their content. Specifically, LLMs evaluate the semantics of webpage content, identifying the subject matter and themes within the text. By measuring the semantic distance or similarity between webpages, LLMs can group related documents, enhancing the search engine's ability to retrieve topically relevant results. This semantic analysis facilitates the automatic labeling and categorization of web pages~\cite{tornberg2024best}. LLMs can analyze the content and context of a webpage, assigning it to appropriate categories or labels based on its semantic characteristics. This process not only streamlines the indexing but also improves the user experience by enabling more accurate and thematic search results~\cite{wu2024well}.

\subsubsection{Query Candidates Generation for Indexing}
LLMs can play a critical role in training neural information retrievers and in the cold start phase of a search engine by generating a list of potential queries related to the content of a webpage. This approach ensures that the search engine is primed with relevant queries for new or less-indexed content, in following two steps.
\begin{itemize}
    \item \textbf{Query Candidate Generation from Contents}: By comprehensively analyzing the content of a webpage, LLMs can generate a list of potential queries that users might input when searching for similar information~\cite{zhou2018neural}. This capability is essential in deciphering the context and intention behind user inquiries, thereby aligning the responses generated by the search engine more accurately with user anticipations. The generated queries provide valuable training data for neural information retrievers~\cite{yuan2023selecting}.
    
    \item \textbf{Cold Start New Contents with Generated Candidates}: By simulating real-user queries, LLMs can help these models learn to predict and rank relevant web pages more effectively, even in scenarios where direct user query data may be limited. For new or niche content that may not yet have associated user queries, the list of generated queries can kickstart the search engine's understanding and indexing of such content. This alleviates the ``cold start'' problem, ensuring that all content, regardless of its current popularity or visibility, can be discovered and retrieved by users~\cite{gong2023unified,huang2024large}.
\end{itemize}
One could use similar prompts in Fig.~\ref{fig:extract-term} to generate candidate queries from the content of a webpage.

\subsection{Augmented~Information Retrieval,~Document Ranking,~and~Content Recommendation}
LLMs have demonstrated remarkable potential in improving the functionalities of search engines, particularly in the area of information retrieval (IR), webpage ranking, and content recommendation as shown below in the next few subsections. 

\subsubsection{Annotation for Retrieval and Ranking}
One of the fundamental challenges in training neural networks for information retrieval lies in the necessity to accurately annotate the relevance of query-webpage pairs~\cite{alonso2008crowdsourcing,le2010ensuring}. This approach involves LLMs in the annotation process for LTR~\cite{chiang2023closer} from three aspects as follows.
\begin{itemize}
    \item \textbf{Point-wise LTR Annotation}: LLMs can assign ranking scores to individual documents relative to a query, based on relevance and user context. These point scores serve as training data for models that aim to replicate such scoring~\cite{werner2022review}. 
    \item \textbf{Pair-wise LTR Annotation}: For pair-wise approaches, LLMs can determine the relative order between any two webpages in response to a query, considering both content relevance and user-specific information. This relative ranking aids in training algorithms to understand preferences within sets of documents~\cite{cao2007learning}. 
    \item \textbf{List-wise LTR Annotation}: In a more comprehensive capacity, LLMs can generate ranked lists of webpages based on their collective relevance and personalization for a query. This ranked order provides a template for list-wise LTR models to learn how to sequence document sets effectively~\cite{xia2008listwise}.
\end{itemize}
Fig~\ref{fig:ltr-llm} illustrates an example of prompts and responses for Point-wise, Pair-wise, and List-wise LTR annotations, where the LLM predicts the relevance score of every retrieved result, the partial order of every two retrieved results, and the order of all retrieved results, subject to the query. By providing high-quality, relevance-annotated pairs, LLMs ensure that the training data for information retrieval neural networks is both accurate and representative of diverse query intents and informational needs~\cite{maoro2023leveraging,saad2023ares,zhao2024dense}.

\begin{figure}
    \centering
    \subfloat[Point-wise LTR Annotation]{\includegraphics[width=0.5\textwidth]{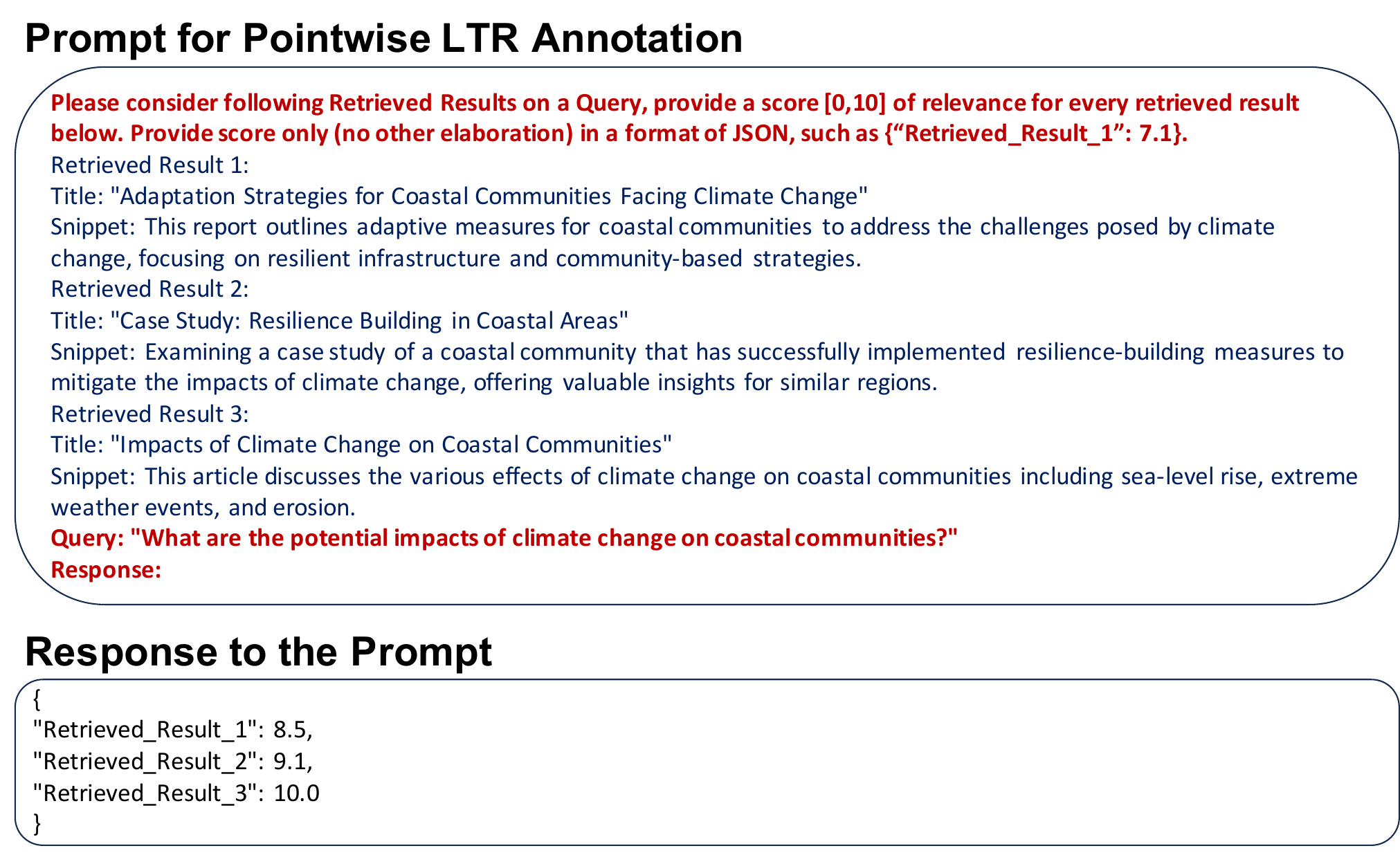}}\\ \vspace{-3mm}
    \subfloat[Pair-wise LTR Annotation]{\includegraphics[width=0.5\textwidth]{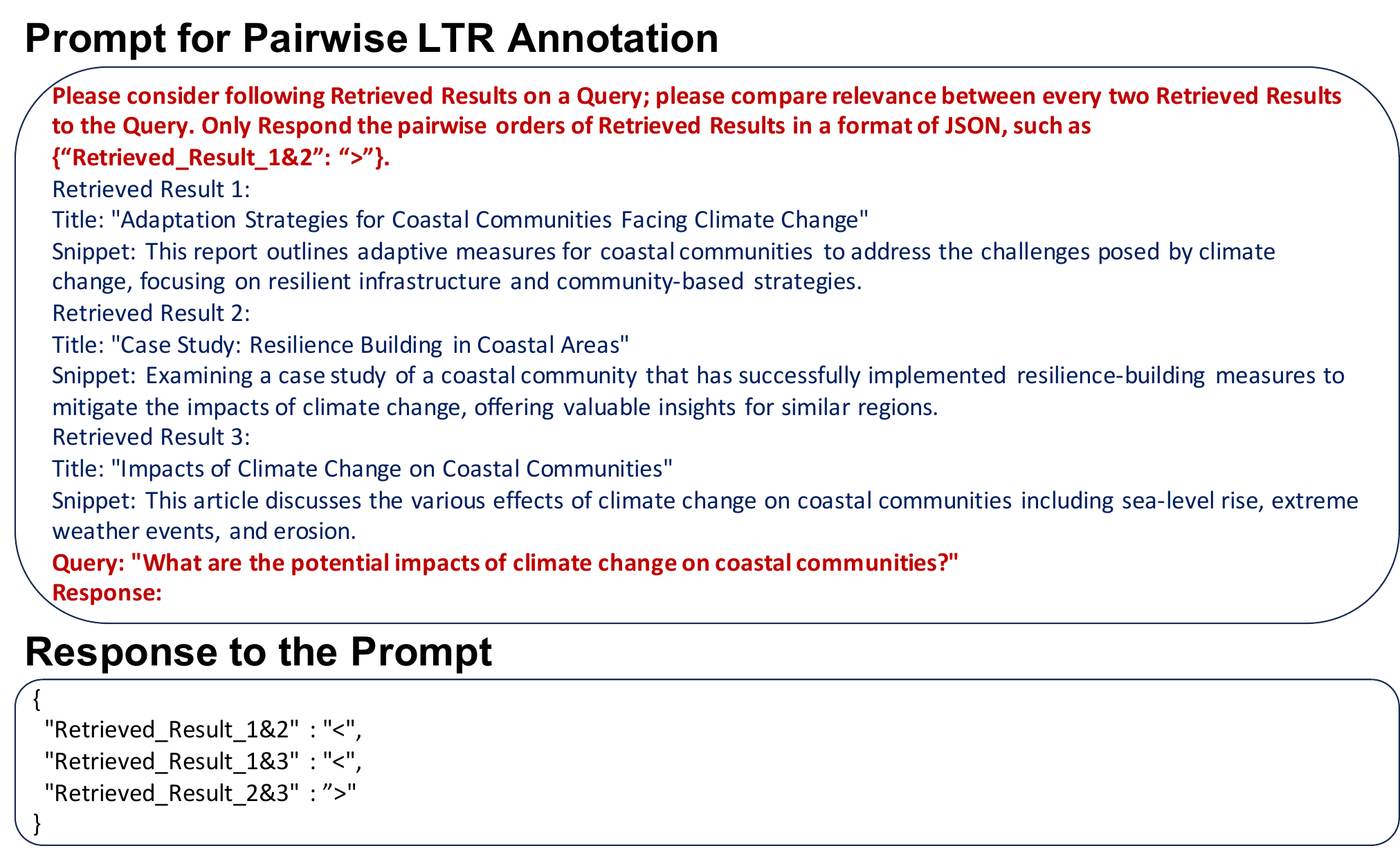}}\\ \vspace{-3mm}
    \subfloat[List-wise LTR Annotation]{\includegraphics[width=0.5\textwidth]{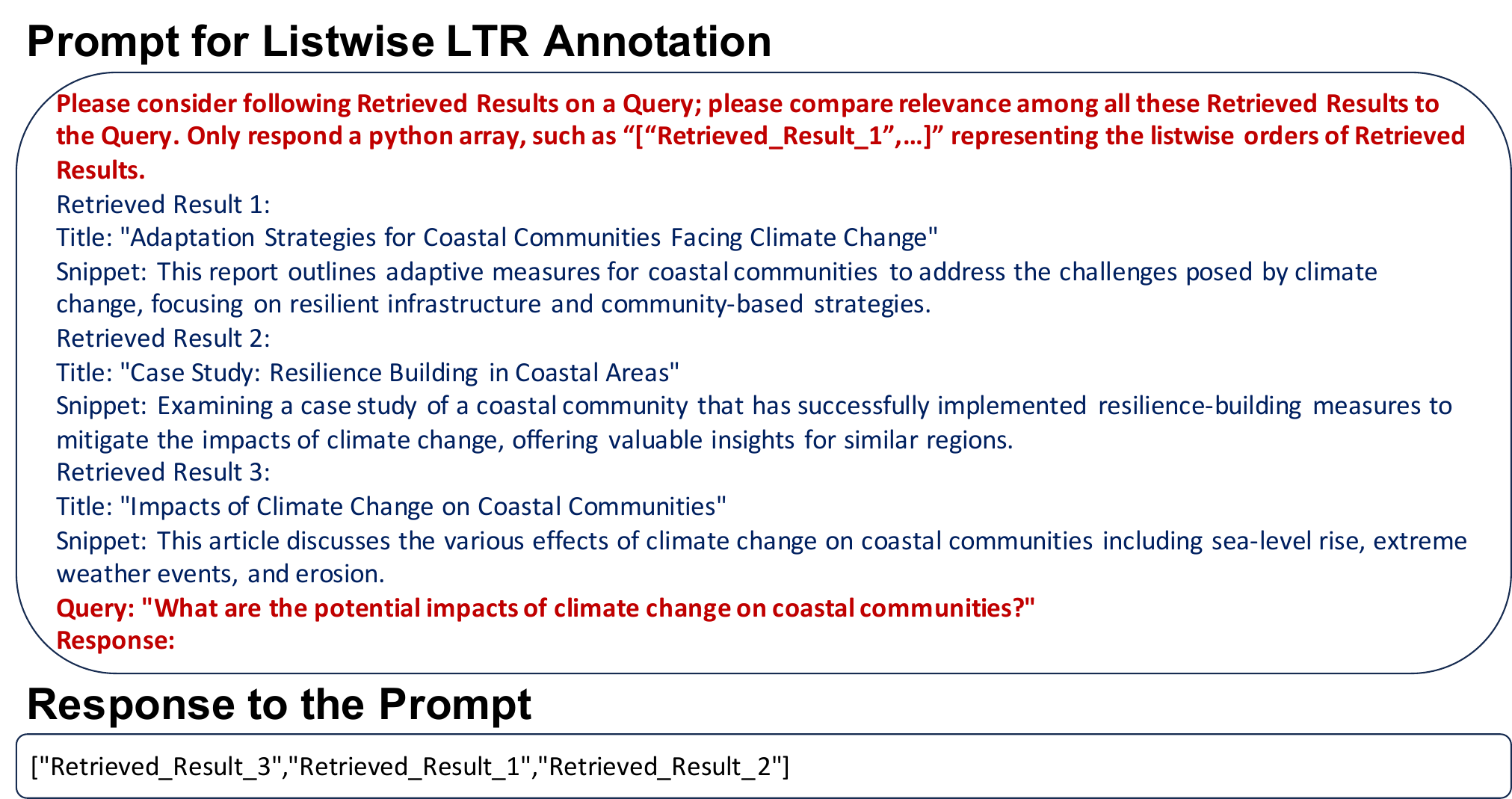}}\vspace{-3mm}
    \caption{An Example of Prompts and Responses for Annotation of Point-wise, Pair-wise, and List-wise LTR}\vspace{-5mm}
    \label{fig:ltr-llm}
\end{figure}

\subsubsection{Online Ranking and Recommendation for Contextual, Personalized Search}
Upon the establishment of an information retrieval model, LLMs can further enhance the search experience by leveraging users' browsing/searching history and profiles to perform online ranking of retrieved webpages or the recommendations of contents. 
\begin{itemize}
    \item \textbf{Ranking}: Given a set of retrieved webpages for a search query, LLMs can evaluate the contextual relevance of each retrieved webpage by considering the specific needs and interests of the user as reflected in their search history and profile. By comparing relevance scores and incorporating personalization factors, LLMs can dynamically adjust the ranking of retrieved webpages, ensuring that the most relevant and personalized results are prioritized for the user~\cite{sun2023chatgpt,zhao2024dense}. Note that one can incorporate similar prompt in Fig.~\ref{fig:ltr-llm} while adding the user's browsing history or profiles as part of contexts in the prompt for enabling online ranking with contextual personalization.
    
    \item \textbf{Recommendation}: In addition to ranking the retrieved webpages for search, yet another way is directly recommend content that the user might be interested in. LLMs can analyze the textual data in user profiles and browsing history, which may include user preferences, demographic information, and real-time interests. For example, during a search session, if a user is looking at sports equipment, the search engine would probably recommend sports-related content or products. To achieve the goal, the so-called \emph{LLM4Rec} techniques have been proposed with LLMs and prompts~\cite{wu2023survey}, where LLMs could be pre-trained and/or fine-tuned to understand users, items in texts~\cite{gong2023unified} and predict the user-item interactions~\cite{wang2024llm} accordingly.

\end{itemize}

\subsubsection{Retrieval-Augmented Generation (RAG) Contents for Conversational Search}
The incorporation of Retriever-Augmented Generation (RAG) into search engines significantly enriches the result output from a generation perspective. Once relevant documents have been retrieved and ranked appropriately, RAG leverages the wealth of factual information within these sources to generate coherent, contextually relevant, and information-rich content for users. Rather than simply returning a list of documents, RAG synthesizes the information extracted from the top-ranking sources to compose responses that effectively combine the retrieved knowledge into a cohesive answer.

\begin{figure}
    \centering
    \includegraphics[width=0.5\textwidth]{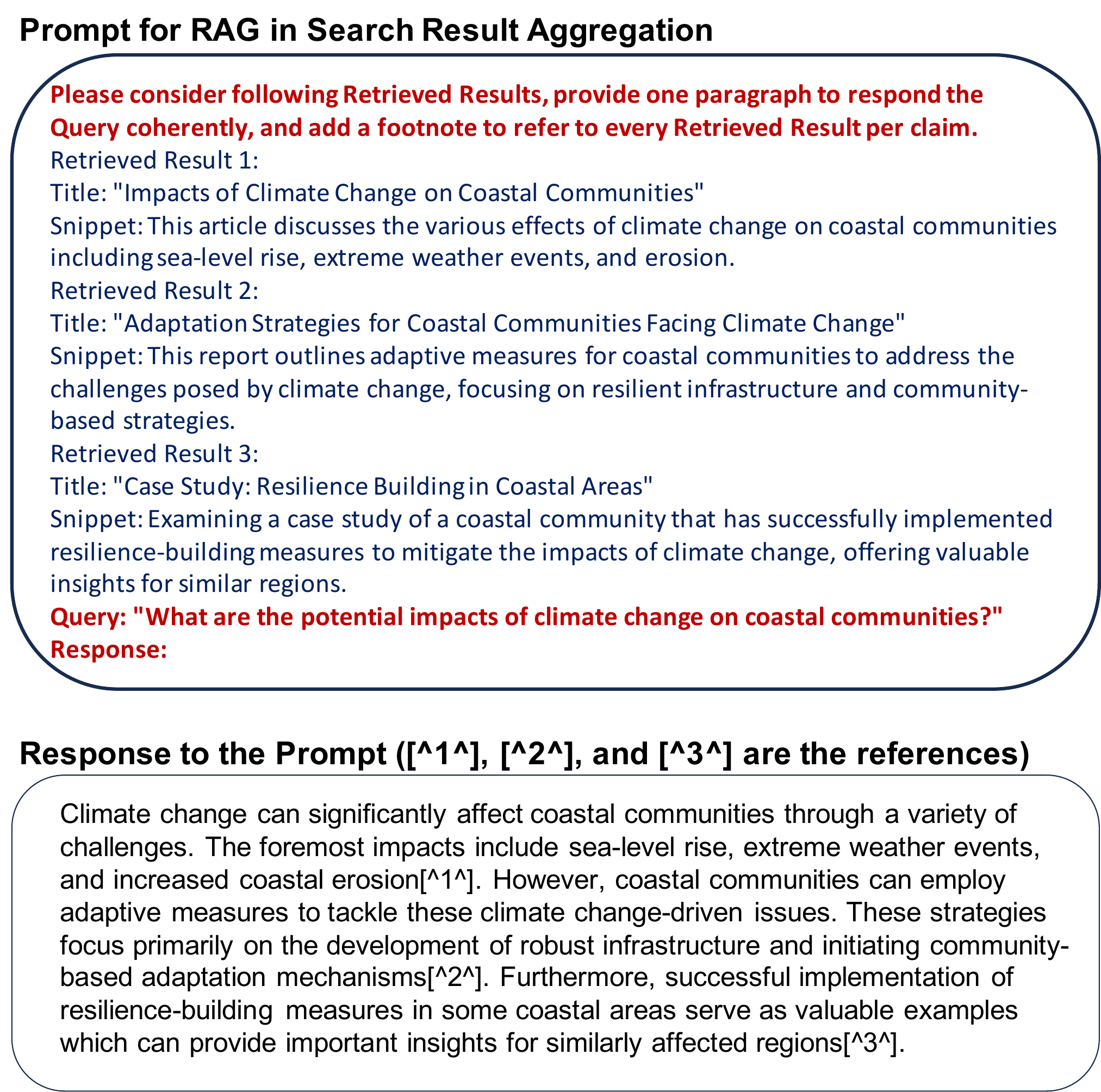}
    \caption{An Example of Prompts and Responses for RAG in Search Results Aggregation}
    \label{fig:search-rag}\vspace{-5mm}
\end{figure}

As shown in Fig.~\ref{fig:search-rag}, given the retrieved results ranked in an appropriate order, RAG could synthesize a coherent response (with a summary, details, and references) to the query through simple prompting with LLMs. From a generation standpoint, the benefits of RAG in search engines may include follows.
\begin{itemize}
    \item \textbf{Enhanced Accuracy and Relevance}: By drawing directly from the retrieved high-quality documents, RAG ensures that the generated responses not only accurately reflect the content of these sources but also maintain a high degree of relevance to the original query.

    \item \textbf{Overall Coherence}: RAG models understand the broader context obtained from the entirety of the retrieved documents, allowing for the production of coherent responses that consider multiple facets of a user's question.

    \item \textbf{Efficient Summary Generation}: They can effectively summarize and condense information from multiple documents, distilling complex data into digestible and accessible formats for the end-user.

    \item \textbf{Data-Rich Responses}: RAG-enabled search engines provide detailed, well-informed answers by cross-referencing various sources, leading to a richer informational value compared to search engines that only offer links.

    \item \textbf{Natural Language Output}: Leveraging the NLP capabilities of underlying language models, RAG produces answers in a conversational tone, which can improve user engagement and understanding.

\end{itemize}
By combining the robust data retrieval aspect with the advanced natural language generation capabilities of GenAI, RAG transforms search engines into powerful tools that don't just find information--they also present it in a instantly usable way, making the search process more seamless and the results more actionable for the user.

\subsection{Augmented Evaluation for Search Engines}
The evolution of search engine technology necessitates equally advanced methods for evaluating performance and user experience. LLMs offer significant potential in augmenting the evaluation of search engines through several innovative approaches, as shown below. 
\subsubsection{Automated A/B Testing through User Mimicking}
LLMs can enhance the efficiency and effectiveness of A/B testing in search engines by acting as agents that mimic user search behaviors. This application allows the direct comparison of different search result sets and their respective ranking orders. Some key features of LLMs are as follows. 
\begin{itemize}
    \item \textbf{Traffic Mockup}: By generating a diverse range of user queries based on real-world search patterns and intentions, LLMs can simulate the natural variability in search behaviors~\cite{huang2024large,yuan2023selecting,zhou2023unified}. 
    \item \textbf{Automatic Evaluation}: LLMs can evaluate two sets of search results (from A/B variants) for the same query, comparing not just the relevance but the ranking order, to gauge which set is more likely to satisfy the user's needs~\cite{chiang2023closer,sun2023chatgpt,saad2023ares}.
    \item \textbf{User Mimicking}: Apart from evaluating results, LLMs can mimic user behaviors in interacting with these results, including clicking through links according to the perceived relevance, thus offering deeper insights into the effectiveness of ranking algorithms~\cite{khandelwal2023large,bernard2024leveraging}.
\end{itemize}
Fig.~\ref{fig:search-compare} illustrates an example of automatic evaluation that compares the two sets of search results under the same query, from the perspectives of relevance, timeness, and the ranking order. In this example, through encapsulating the titles and snippets of webpages in the order of search results into the prompt, GPT-4 could respond the evaluation results automatically from the perspectives desired, and formates the result in a programming-friendly way. Actually, GPT-4 can also generate interpretations on the comparison. Due to the page limit, we haven't include the full response here.
\begin{figure}
    \centering
    \includegraphics[width=0.5\textwidth]{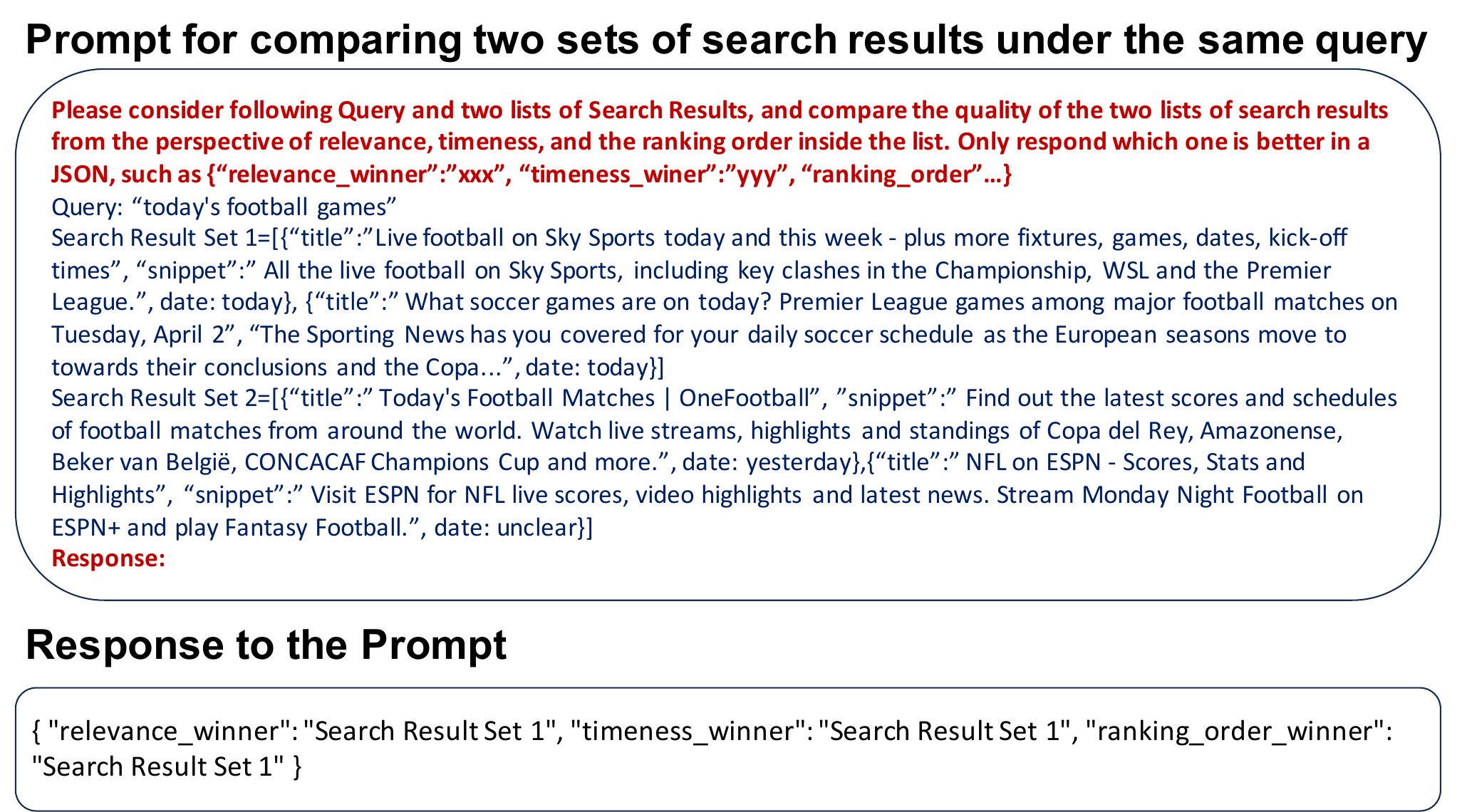}
    \caption{An Example of Prompts and Responses for Automatic Evaluation of Search Results}
    \label{fig:search-compare}
    \vspace{-5mm}
\end{figure}

\subsubsection{Decoding User Interactions and Intentions}
Through text understanding, LLMs can interpret the user interactions with search engines. This capability allows for a deep understanding of user satisfaction and intent changes throughout their search trajectories.  Some key features of LLMs are as follows. 
\begin{itemize}
    \item \textbf{Sequential Behavior Modeling}: LLMs can analyze patterns in click-throughs and the order of interactions to infer the relevance and quality of the search results provided. By examining the changes in user queries during a search session, LLMs can infer shifts in user intentions or pinpoint when a user's needs become more specific~\cite{wang2023drdt,lin2023rella,khandelwal2023large}. 
    \item \textbf{Satisfaction Assessment}: The change in these sequential behaviors can signal how well the search engine accommodates evolving user needs. LLMs can assess whether the search engine helps users find desired information with minimal steps of user interactions (e.g., clicks or queries), indicating the efficiency and effectiveness of the search process~\cite{khandelwal2023large,lin2024interpretable}.
\end{itemize}

\subsubsection{Evaluation with User Experience Dashboards}
LLMs can play a vital role in transforming raw evaluation data, including A/B testing outcomes and user interaction analyses, into comprehensive dashboards that highlight key aspects of user experience, as follows. 
\begin{itemize}
    \item \textbf{Data Aggregation}: By aggregating data from numerous comparisons and user interactions, LLMs can pinpoint critical performance metrics such as click-through rates, query refinement patterns, and satisfaction indicators~\cite{lewis2020retrieval,hosseini2022knowledge}. 
    \item \textbf{Data Interpreter}: Advanced data synthesis capabilities enable LLMs to identify patterns and trends suggesting areas for improvement, whether in search algorithms, result ranking, or user interface design~\cite{hong2024data,liu2024llms}. 
    \item \textbf{Summary \& Reporting}: Leveraging their language generation capabilities, LLMs can generate insightful, understandable narratives around the data, complemented by visual dashboards that highlight search engine performance from the user's perspective~\cite{zhao2023new}. 
\end{itemize}
These dashboards can serve as invaluable resources for developers and designers looking to enhance search engine technologies.

\subsection{Summary of LLM4Search Research}
Incorporating LLMs with search engines heralds a pivotal shift in the paradigm of information retrieval, query processing, and user interaction. These advanced models offer a suite of capabilities that significantly enhance the efficiency, accuracy, and user experience of search engines. Upon examining their multifaceted contributions, it becomes evident that LLMs possess  potential by four key abilities as follows. 
\begin{itemize}
    \item \textbf{Content Understanding and Information Extraction}: LLMs exhibit an unparalleled proficiency in dissecting and interpreting search queries, web content, browsing history, and user profiles. They adeptly extract relevant information by understanding the semantic meaning contained within queries and documents. This deep understanding enables LLMs to parse web pages for precise index term extraction, categorize content semantically, and tailor query suggestions based on historical user data. Adopting this capability ensures that search engines can process queries with a higher degree of accuracy, leading to improved retrieval that aligns closely with user intent.

    \item \textbf{Semantic Relevance for Content Matching and Ranking}: At the heart of LLMs' functionality is their ability to analyze and evaluate semantic relevance, allowing for more advanced content matching and ranking algorithms. LLMs leverage their extensive knowledge base and language understanding to match queries with the most relevant content, even when the match is not immediately apparent from the query's keywords alone. This semantic analysis extends to the generation of contextual queries and enhances the search engine's ability to categorize web pages, contributing to a more comprehensive understanding of content relevance and significantly improving the quality of search results.

    \item \textbf{User Profiling and Context Modeling}: LLMs stand out for their capability to offer highly contextualized and personalized search experiences through in-context learning. By analyzing current search queries in conjunction with historical user data, LLMs can craft responses that are tailored to the individual's specific needs, preferences, and patterns of behavior. This level of personalization not only enhances user satisfaction but also makes information retrieval more efficient by prioritizing results that are most relevant to the user's context and search history.

    \item \textbf{Comparative Analysis for Ranking and Evaluation}: Finally, LLMs excel in their ability to conduct comparative analyses, whether for the purpose of webpage ranking or for evaluating the effectiveness of search results. Through automated relevance annotation, contextually personalized online ranking, and annotating capabilities for learning-to-rank tasks, LLMs can significantly refine webpage ranking processes. Additionally, their role in automating A/B testing and synthesizing user interaction data into actionable insights marks a significant advancement in search engine evaluation. This ability to dynamically adapt and refine ranking parameters ensures that search engines can continuously evolve to meet and exceed user expectations with a high degree of precision.

\end{itemize}
As technology continues to advance, the partnership between LLMs and search engines will undoubtedly lead to even more innovative solutions that will shape the future of how we interact with information.  

\section{Challenges and Future Directions}
Search engines - LLMs symbiosis under the research themes of \emph{Search4LLM} and \emph{LLM4Search} is very promising. However, there are many technical challenges that must be overcome. We discuss some of these challenges and future research directions along these lines.

\subsection{Memory-Decomposable LLMs}
Efficiently managing and updating the extensive memory stores of LLMs, whether for enhancing search engines with LLMs or vice versa, poses a significant challenge in delivering accurate, context-aware responses in real-time~\cite{meng2022mass,mitchell2022memory}. We identify several technical issues as follows.
\begin{itemize}
    \item \textbf{Memory at Scale}: The scalability of CRUD operations, including creation, read, update, and detection, within the memory components of LLMs is critical to their effective functioning. 
    
    \item \textbf{Consistency and Integrity}: Maintaining data consistency and integrity during CRUD operations in LLMs is challenging, especially when dealing with real-time data updates and deletions, which could render part of the model's knowledge obsolete or incorrect.
    
    \item \textbf{Support to Efficient Retrieval and Editing}: Ensuring that the LLM can accurately understand and utilize its decomposed memory segments for CRUD operations to maintain contextual relevance and coherence in responses. This involves sophisticated understanding and integration of user queries and stored knowledge.
\end{itemize}
Considering the technical challenges outlined above, it may be worthwhile to explore the following research directions:
\begin{itemize}
    \item \textbf{Novel Architecture}: Developing more advanced memory management algorithms and architectures that permit LLMs to more effectively access and update their knowledge base. This design could involve techniques like dynamic memory allocation or memory networks that can selectively store and retrieve information.

    \item \textbf{Incremental Learning}: Introducing techniques for incremental learning that allow LLMs to update their memory with new information efficiently, or even forget specific piece of (outdated) information, are crucial for implementing CRUD operations effectively.
    
    \item \textbf{Exact Retrieval by Generation}: Current LLMs' responses are generated through predicting token sequences with maximal probabilities. These responses are often plausible but incorrect information, a phenomenon known as ``hallucination''~\cite{li2023helma}. However, in search engine settings, a memory-decomposable LLM  might need to perform information retrieval from all data it has traversed during pre-training or fine-tuning. Thus, there is a need to innovate a method for ``exact-recovery'' of (part of) the training corpus. 
\end{itemize}

\subsection{Explainability of LLMs for Web Search}
LLMs often operate as ``black boxes'', making it difficult to understand how they arrive at their outputs. This lack of explainability can be problematic when using LLMs to augment search engines, as it may be challenging to interpret or trust the results~\cite{li2022interpretable,xiong2024towards}.     

\begin{itemize}
    \item \textbf{Model Complexity}:  LLMs are usually built on over-parameterized architectures with hundreds of millions, or even billions, of parameters. This complexity makes it difficult to pinpoint the exact reasoning behind a given output. The inability to understand the inner workings complicates the task of ensuring the reliability and trustworthiness of LLMs in query understanding, rewriting and so on~\cite{zhao2023explainability}.

    \item \textbf{Opaque Decision-Making Process}: LLMs provide limited insight into their decision-making process. This opaqueness hinders the ability to identify the source of errors or biases in the model's outputs. When LLMs are used to augment search engines (LLM4Search), or when search engines are used to improve the performance of LLMs (Search4LLM), the lack of transparency can erode user trust in the search results~\cite{li2022interpretable}.

    \item \textbf{Scale of Data}: LLMs are trained on massive datasets. Tracing back the influence of specific data points on the output becomes practically impossible. The use of web-scale datasets scales up the challenge in understanding which part of the data contributed to any misinformation or biased information being relayed through the model~\cite{NEURIPS2022_d0702278,xiong2024towards}.
\end{itemize}
Addressing the explainability challenges of LLMs, under the themes of \emph{Search4LLM} and \emph{LLM4Search}, is critical for advancing the trustworthiness and utility of these technologies. Future research would need to balance the trade-offs between explainability, accuracy, and efficiency to build more transparent, accountable, and user-friendly LLMs. Some promising directions are as follows.
\begin{itemize}
    \item \textbf{Development of Interpretable Models}: There's a pressing need for research on creating LLM architectures and training methodologies that inherently support explainability. This includes developing models that can articulate the reasoning behind their responses in a manner that is understandable to users.

    \item \textbf{Explainable AI (XAI) Techniques for LLMs}: Investigating and adapting XAI techniques to the specific context of LLM4Search and Search4LLM could offer insights into how these models process and retrieve information. This includes creating visualization tools and summary techniques that can help demystify the model's internal processes~\cite{zhao2023explainability}.

    \item \textbf{Bias and Fairness in Model Explanations}: Ensuring that explanations not propagate biases present in training data or amplify unfair representations requires dedicated research. This could involve developing methods to audit and refine model explanations for equity and inclusiveness~\cite{li2022interpretable}.

    \item \textbf{User-Centric Explanation Frameworks}: Developing frameworks that can adapt explanations based on the user's background knowledge and the context of the query. This could involve personalized explanation systems that adjust the complexity and detail of explanations accordingly.

\end{itemize}

\subsection{Agents for Search4LLM and LLM4Search}
Both themes request LLMs being able to understand queries and contents while working with other components to fulfill the goals. As was defined in~\cite{li2023autonomous}, an agent is built upon the capabilities of memory, planning, and action with LLMs. Thus, to enable agents for \emph{Search4LLM} and \emph{LLM4Search}, some technical challenges should be addressed as follows.
\begin{itemize}
    \item \textbf{Integration Complexity}: Agents need to be seamlessly integrated with the vast and complex submodules, components, and tools of LLMs and search engines. This includes the capability to access and interpret vast datasets, understand context from partial or ambiguous queries, and manage real-time data fetching without compromising response times~\cite{reed2022a,xiong2023natural}.

    \item \textbf{Long/Short-Term Memory for Interactions}: For agents to effectively contribute to both \emph{LLM4Search} and \emph{Search4LLM}, they must possess sophisticated memory capabilities. This involves not only storing and retrieving information but understanding the relevance of historical interactions in current contexts. How they adapt their memory systems for dynamic and efficient use is a key challenge~\cite{hatalis2023memory}.

    \item \textbf{Adaptive Planning}: Agents must plan their actions in environments that are constantly evolving. In the context of \emph{LLM4Search} and \emph{Search4LLM}, this issue requests adapting to changes in user behavior, search patterns, and the availability of online content. Planning in such an adaptive manner requires continuous learning and adjustment mechanisms~\cite{wang2023describe}.

    \item \textbf{Action and Interaction}: Taking appropriate actions based on the contextual understanding and planning involves interacting with both internal systems of LLMs and external tools (e.g., crawlers, databases, and search engines). Ensuring these actions are both relevant and timely, while minimizing errors or irrelevant outputs, is challenging~\cite{reed2022a,deng2024mind2web}.


\end{itemize}
To address above challenges, promising directions for future research are as follows.
\begin{itemize}
    \item \textbf{Improved Memory Space}: Developing advanced memory spaces that allow agents to more effectively store, retrieve, and utilize knowledge over time. This could involve exploring neuromorphic computing models or advanced neural network designs that mimic human memory processes more closely, or leveraging transformer models that can handle extreme/infinite length of context to restore all previous interactions by texts.
    
    \item \textbf{Dynamic Planning Algorithms}: Researching algorithms that enable more flexible and dynamic planning based on real-time data and changing environments. This could include reinforcement learning approaches that adapt based on success/failure feedback loops.
    
    \item \textbf{Interactive Learning Models}: Developing models that allow agents to learn from their interactions not just with users but with other AI systems and online databases. This approach could lead to more comprehensive understanding and action-taking abilities.
    
    \item \textbf{Cross-domain Knowledge Transfer}: Exploring methods for more effective cross-domain knowledge transfer and application. This involves agents not just specializing in one area but being able to apply insights from one domain to another fluently.
    
    \item \textbf{Real-time Data Processing and Action Taking}: As the need for immediate and pertinent information grows, investigations into how agents can manage real-time data and make prompt decisions without compromising precision or relevance becomes a focus area.  
\end{itemize}

\subsection{Miscellaneous}
In addition to above well-structured discussions, some miscellaneous technical challenges and promising research directions are as follows.
\begin{itemize}
    \item \textbf{Data Quality and Bias}: Ensuring the accuracy and fairness of information retrieved or utilized by LLMs. The inherent biases in training data can skew search results or LLM responses, potentially propagating misinformation.

    \item \textbf{User Satisfaction and Trust}: Building and maintaining user trust in the accuracy and reliability of LLM-augmented search engines. Users might be skeptical about algorithmic transparency and the quality of personalized results. 

    \item \textbf{Intellectual Property and Privacy Concerns}: Using content from the web to train LLMs raises significant concerns over copyright infringement and personal data privacy. 
    
    \item \textbf{Legal and Ethical Considerations}: Navigating the complex landscape of regulations governing AI and data use across different jurisdictions. The use of LLMs in decision-making processes further complicates this, requiring ethical and responsible AI systems. 

\end{itemize}
With respect to above challenges, some promising research directions are as follows.
\begin{itemize}
    \item \textbf{Enhanced Methods for Bias Detection and Correction}: Innovating more sophisticated AI algorithms that can detect various forms of bias in data and automatically correct or mitigate these biases.

    \item \textbf{User-Centric Design and Feedback Mechanisms}: Implementing design principles that put the user first, including customizable privacy settings and the introduction of mechanisms where users can provide real-time feedback on the relevance and quality of search results.

    \item \textbf{Cross-Disciplinary Research on Legal and Ethical AI Use}: Conducting cross-disciplinary research that involves legal scholars, ethicists, technologists, and policymakers to develop standards and guidelines for the ethical use of AI in search and information retrieval.

\end{itemize}
By focusing on these challenges and research avenues, the development of \emph{LLM4Search} and \emph{Search4LLM} can be guided toward more cost-effective, responsible, user-friendly, and legally compliant systems that leverage the strengths of LLMs while addressing their inherent limitations.

\section{Discussions and Conclusions}
This work endeavors to elucidate the reciprocal relationship between Large Language Models (LLMs) and search engines, dissecting how each entity could potentially enrich and augment the functionalities of the other. 

\subsection{Core Visions}
Under the \emph{Search4LLM} umbrella, the focus is placed on how the vast, diverse datasets available from search engines could be harnessed to enhance the pre-training and fine-tuning processes of LLMs. This approach aims at bolstering the LLMs' grasp on query contexts thereby elevating their precision in generating responses that are both relevant and accurate. The premise hinges on utilizing high-quality, ranked documents as prime training data, underlining the significance of such data in improving  overall understanding and response generation capabilities of LLMs. The exploration into Learning To Rank (LTR) algorithms further underscores an attempt to refine abilities of LLMs in analyzing and prioritizing information relevance, essentially sharpening their effectiveness in response accuracy and relevancy.

In contrast, the theme \emph{LLM4Search} shifts focus to examine the potential influence that Latent Language Models (LLMs) may impart on improving the functional aspects of search engines. Here, the narrative shifts to leveraging LLMs for tasks like effective content summarization, aiding in the indexing process, and employing fine-grained query optimization techniques to yield superior search outcomes. Moreover, the role of LLMs in analyzing document relevance for ranking purposes and facilitating data annotation in various LTR frameworks is underscored. This segment hints at a realm where LLMs do not just passively benefit from search engine data but actively contribute to improving the efficiency, accuracy, and user experience of search engine platforms.

\subsection{Challenges and Opportunities}
In conclusion, the intersection of LLMs and search engine technologies presents a fertile ground for innovation, offering avenues to transcend current limitations in both domains. The \emph{Search4LLM} initiative underscores the rich potential that search engine datasets have in refining the operational intelligence of LLMs, enabling these models to more adeptly handle query complexities—a leap towards smarter, more adaptive, and user-centric search services. Meanwhile, \emph{LLM4Search} showcases the transformative impact that LLMs could have on the search engine ecosystem, enhancing content understanding, search precision, and user satisfaction.

However, the path to fully integrating LLMs with search engines is fraught with challenges, including technical implementation hurdles, ethical considerations, biases in model training, and the need to keep training datasets current with the evolving internet landscape. Despite these challenges, this work illustrates a promising horizon where the synergistic marriage between LLMs and search engines could herald a new era of intelligent, efficient, and user-centric search services. This exploration not only contributes to the advancement of services computing but also lays a systematic framework for future research and development in this dynamic intersection of technologies.  

\bibliographystyle{abbrv}
\bibliography{main}

\end{document}